\documentclass[reprint,amsmath,amssymb,aps,prb,superscriptaddress,floatfix,nofootinbib]{revtex4-2}
\usepackage{graphicx}
\usepackage{siunitx}
\usepackage[T1]{fontenc}
\usepackage[utf8]{inputenc}
\usepackage{xcolor}
\usepackage{lmodern}
\definecolor{codegreen}{rgb}{0,0.6,0}
\definecolor{codegray}{rgb}{0.5,0.5,0.5}
\definecolor{codepurple}{rgb}{0.58,0,0.82}
\definecolor{backcolour}{rgb}{0.95,0.95,0.92}
\definecolor{urlblue}{HTML}{007bff}

\usepackage{bm}
\usepackage{bbm}
\usepackage{orcidlink}
\usepackage{microtype}
\usepackage{enumitem}
\usepackage{braket}
\usepackage{mathtools}
\usepackage{soul}
\usepackage[normalem]{ulem}
\usepackage{nicefrac}
\usepackage{lipsum}
\usepackage{slashed}

\usepackage{newtxtext,newtxmath}

\usepackage[capitalise]{cleveref}
\input{macros.sty}

\hypersetup{
 colorlinks = true,
 linkcolor = urlblue,
 citecolor = urlblue,
 urlcolor = urlblue,
}

\usepackage[mathlines]{lineno}

\begin{document}

\title{
Simple invariants for band topology 
}

\author{Adam Yanis Chaou\orcidlink{0000-0002-9926-4633}}
\affiliation{Donostia International Physics Center, P. Manuel de Lardizabal 4, 20018 Donostia-San Sebastian, Spain}
\author{Adolfo G. Grushin\orcidlink{0000-0001-7678-7100}}
\affiliation{Donostia International Physics Center, P. Manuel de Lardizabal 4, 20018 Donostia-San Sebastian, Spain}
\affiliation{IKERBASQUE, Basque Foundation for Science, Maria Diaz de Haro 3, 48013 Bilbao, Spain}
\affiliation{\small Universit\'e Grenoble Alpes, CNRS, Grenoble INP, Institut N\'eel, 38000 Grenoble, France}
\author{Peru d'Ornellas\orcidlink{0000-0002-2349-0044}}
\affiliation{Donostia International Physics Center, P. Manuel de Lardizabal 4, 20018 Donostia-San Sebastian, Spain}

\date{\today}

\begin{abstract}
Despite the exhaustive understanding gathered around non-interacting topological states of matter, there is no single method capable of systematically delivering simple, numerically efficient topological invariants that is applicable to all crystalline and non-crystalline systems alike. Here we revisit the spectral localizer operator, constructed from the Hamiltonian and position operators, and show how it can be treated it as an auxiliary zero-dimensional Hamiltonian whose topology encodes the higher dimensional phases of the parent Hamiltonian. Its classification reduces every topological invariant to a matrix signature or the sign of a Pfaffian for an appropriate localizer, both of which are simple to interpret and efficient to compute in real space. We validate this approach by deriving simple real-space invariants for weak and rotationally invariant crystalline phases that were previously beyond the grasp of the spectral localizer formalism, atomic limits that escape scattering invariants, phases that evade symmetry-based indicator methods, as well as phases that had no previously known invariant. Our work provides a systematic way to construct any non-interacting topological invariant for a crystalline or non-crystalline systems, opening avenues to classify and predict the topology of previously unexplored classes of materials.
\end{abstract}

\maketitle
\tableofcontents

\section{Introduction}

Our experimental and theoretical understanding of non-interacting topological phases rests on the ability to classify and diagnose them by calculating topological invariants~\cite{Chiu2016}. While the question of classifying topological phases has an exceptionally mature answer thanks to decades of developments in K-theory~\cite{kitaev2009periodic,ryu_topological_2010,freed2013twisted,morimoto2013,chiu2013,shiozaki2014topology,shiozaki2016nonsymmorphic,shiozaki2017crystalline2,trifunovic2017scattering,geier2018second,cornfeld2019,okuma2019pointgroupHOTI,trifunovic2019higher,cornfeld2021,shiozaki2022spacegroups,shiozaki2023generalized}, 
there seems to be no unique strategy to construct explicit, computable, let alone numerically efficient, expressions of topological invariants. 

The problem has many facets. Crystalline phases have been subject to vast K-theoretic classifications which, however, often lack accompanying invariants. 
Translational symmetry allows for the expression of invariants in momentum space~\cite{vanderbilt_berry_2018,Bernevig2013_TI_TS,brouwer2023homotopic}, but this choice excludes disordered systems from the outset. 
Using additional crystalline symmetries aids high-throughput searches over a vast number of crystalline materials ~\cite{Zhang2019,Vergniory2019,Tang2019,Tang2019NatPhys,Vergniory2022} using symmetry-based indicators~\cite{fu2007inversion,Kruthoff2017,Po:2017ci,bradlyn_topological_2017,Song2018,Po2020,Cano2021}, but certain topological states fall outside their scope~\cite{Moore2008,Fang2012,kooi2021bulk,Kooi2019,Bouhon2020,geschner2026band}. 
Additionally, invariants can take complicated mathematical forms, which work against widespread adoption, useful physical insight, and efficient numerical implementation.

Real-space methods are a useful alternative. Scattering invariants~\cite{fulga2012scattering,trifunovic2017scattering,zijderveld2025scattering} succeed because they are efficiently implemented in real space and track topological properties protected by mobility gaps. However, these fail to diagnose atomic-limits, because of the absence of anomalous boundary states, and not all scattering invariants have been formulated for higher-order topological states, even when no obstruction exists~\cite{zijderveld2025scattering}. Moreover, it is not possible in general to write invariants for exact non-crystalline symmetries, like those in quasicrystal space-groups~\cite{Rabson1991}. 
While these and other real-space methods are useful for disordered crystals~\cite{li_topological_2009,groth_theory_2009,Guo2010,Prodan2011,Prodan2011b,Song2012,Ringel2012,Lv2013,fulga_statistical_2014,Ni2020,Silva2024,chaou2025,zijderveld2026symmetric}, 
quasicrystals~\cite{Kraus2012,Tran:2015cj,Bandres2016,Fuchs:2016hp,Huang2018,Huang2018,Fuchs:2018dd,Loring2019,varjas_topological_2019,Huang2019,Chen2019,He2019,Duncan2020,Fan2021,Zilberberg:21,Else2021,Hua2021,Jeon2022,Schirmann2024,Manna2024,rochecarrasco2025,caiger2026fractaltopologymajoranabound,manna_higher_2022,manna_anyons_2020}, amorphous solids~\cite{agarwala_topological_2017,mansha_robust_2017,xiao_photonic_2017,mitchell_amorphous_2018,bourne_non-commutative_2018,poyhonen_amorphous_2018,minarelli_engineering_2019,chern_topological_2019,mano_application_2019,costa_toward_2019,marsal_topological_2020,mukati_topological_2020,sahlberg_topological_2020,ivaki_criticality_2020,agarwala_higher-order_2020,Grushin2020,Huang2020,wang_structural-disorder-induced_2021,focassio_structural_2021,Focassio2021,Mitchell2021,spring_amorphous_2021,wang_structural_2022,Ma2022,Peng2022,uria-alvarez_deep_2022,guzman_geometry_2022, spillage_2022,Cassella2022,Grushin2023,Cheng23,zhang2023anomalous,Manna2024,Li2024,marsal_obstructed_2022, uria2025}, 
fractals~\cite{Marta2018,Manna2022,Manna2024}, or hyperbolic lattices~\cite{Urwyler2022,Liu_ChernHyperbolic_2022,Lenggenhager2022} no material has been identified whose topology is protected solely by a non-crystallographic symmetry.

The above observations outline a challenge: to find a single, systematic formulation for \textit{all} topological invariants of a gapped non-interacting Hamiltonian given its symmetries. Ideally, such invariants should be simple, be as easy as possible to interpret, and be sufficiently numerically efficient to be useful. Writing them in real-space will ensure that not only crystalline momentum-space invariants are included, but that it also incorporates non-crystalline invariants. Real-space formulations would not only allow us to systematically apply them to quasicrystals, where no single \textit{material} has of yet been predicted as topological, but also would allow us to quantify the fate of topology subjected to disorder of any nature: on-site, hopping, or structural~\cite{zallen_physics_1998}. The purpose of this work is to describe such a method.

The strategy we follow is to reduce the problem of finding topological equivalences of $d$-dimensional systems into the problem of finding the topology of a \textit{representative} 0-dimensional system---here a 0D systems is just a matrix. The 0D topological classification is simple, since it reduces to classifying whether two matrices can be deformed into one another while remaining invertible along the way and preserving any desired symmetries.
Because we are considering matrices, the invariants necessarily take simple mathematical forms, given by signatures (the difference between the number of positive and negative eigenvalues), determinants, or Pfaffians. These are numerically efficient to compute by using sparse matrix algorithms.

We develop this strategy using the spectral localizer formalism~\cite{LORING2015,Loring2017finite,loring_2019_guide,schulz2024topological,Cerjan_2024_Tutorial}, which has previously yielded real-space, numerically efficient invariants for all AZ classes~\cite{loring2019spectral,loring2020spectral,doll2021skew,schulz-baldes_spectral_2021,stoiber2024spectrallocalizerapproachstrong,doll2024local,wong2026}, for a subset of crystalline
symmetries~\cite{Cerjan2024crystal}, and for fragile phases~\cite{Lee2025}, and has been extended to gapless and semimetallic systems~\cite{cerjan_local_2022,schulz2022invariants,schulz2023spectral,franca2024topological,Franca2024_PRM},
non-Hermitian~\cite{cerjan2023spectral} and driven~\cite{Ghosh2024floquet} settings, and to photonic platforms~\cite{cerjan_operator-based_2022,Dixon2023gapless,Wong2023nonlinear}.
The spectral localizer is an operator built from the Hamiltonian and position operators, whose spectrum encodes whether $H$ and $X$ can be deformed to commute. If they commute, $H$ and $X$ share common eigenstates, which implies that the state is trivial---often referred to as a trivial atomic limit. In this work we consider the localizer as a 0D representative Hamiltonian that we can classify and deform. This perspective allows us to exhaust the classification of topological invariants within the localizer picture, providing a single recipe to write real-space invariants for each topological class, including those that were not previously possible.


Using this formalism, we are able not only to recast previously existing invariants into novel and simple real-space expressions but also to derive simple expressions for previously unknown invariants. Out of the examples we discuss, we highlight existing momentum-space invariants that have thus far been outside the grasp of the localizer formalism, notably weak invariants and invariants for phases protected by rotational symmetries. We also discuss atomic limits phases, which were previously inaccessible with scattering or localizer invariants. We then provide a much simpler invariant for a phase protected by both $C_4$ and $\mathcal{T}$ symmetries, whose invariant takes an involved mathematical form and that is not captured by symmetry-indicator-based methods~\cite{kooi2021bulk}.

The structure of the paper is as follows. In \cref{sec:zero_dimensions} we provide a review of the topological classification of non-interacting systems in zero dimensions, in the ten Altland-Zirnbauer (AZ) symmetry classes, as well as in the presence of additional symmetries. 

In \cref{sec:dim_recg_and_cliff} we introduce the spectral localizer and show how it encodes both the constraints central to band topology, gapped-ness and locality. We discuss how it can be understood as a dimensional reduction scheme, where the localizer ($\mathcal L$) may be treated effectively as a zero-dimensional system that encodes the topological information of the $D$-dimensional parent Hamiltonian. This allows us to extend the zero-dimensional classification to arbitrary higher-dimensional systems. We show how the each dimension reduction changes the symmetry class of the localizer following the Bott clock. We then apply the spectral localizer to Dirac Hamiltonians describing the low-energy spectrum around a topological phase transition. This completes our review of the existing localizer literature. 

In \cref{sec:crystalline_phases} we generalize this methodology to include arbitrary real-space symmetries beyond the conventional AZ classification. We consider a system with a generic symmetry, which can be commuting or anti-commuting, and unitary or anti-unitary. We show how such a symmetry of the Hamiltonian necessarily implies a symmetry of the localizer. This provides a methodology for using the localizer to classify \textit{non-interacting symmetry protected} topological phases.

Finally, in \cref{sec:examples} we provide a number of examples of topological systems with different, and often relatively involved, topological classifications. We show how each may be straightforwardly classified using the architecture proposed here, validating our method as a universal language for characterising non-interacting topological phases in real space. We start with simple examples, showing how weak, metallic and inversion symmetric materials can be classified, before moving on to rotation-protected phases which even in momentum space only admit extremely complex invariants. In particular we find phases that fall outside the existing momentum-space invariant literature, illustrating that the localizer is not only useful for computing invariants when momentum space in unavailable, but can even classify translationally symmetric phases more easily than momentum space invariants.


\section{Topology in zero dimensions}\label{sec:zero_dimensions}
We start with a recap of the simplest topological classifications---those of non-interacting Hamiltonians in zero dimensions. Here a quantum system is simply represented by a Hermitian Hamiltonian matrix, with no real-space coordinates to consider whatsoever. 

The central question here is: given two gapped Hamiltonians $H$ and $H'$, does there exist a continuous path 
\begin{align}\label{eqn:homotopy}
    H(\lambda) \textup{ for }\lambda \in [0,1],
\end{align}
where $H(0) = H$, $H(1) = H'$, such that $H(\lambda)$ remains invertible (i.e.~gapped spectrum) for all $\lambda$. 
If such an interpolation exists then the two Hamiltonians are said to be topologically equivalent. If not, they're said to be in different topological phases. The aim is to construct a topological invariant that can distinguish these phases. 

In addition to remaining gapped along the entire interpolation, we can further require that a symmetry be respected throughout. We shall consider four types of symmetry. The first three: chiral symmetry, particle-hole symmetry (PHS) and time reversal symmetry (TRS) locate the system in one of the Altland-Zirnbauer symmetry classes \cite{zirnbauer_riemannian_1996,altland_nonstandard_1997}. Finally, we shall provide a review of how to adapt the classification in the presence of conventional unitary symmetries. As this material is well-studied this section serves as a self-contained review before applying this construction to higher-dimensional topological phases.

\subsection{Altland-Zirnbauer symmetry classes}
\label{sec:AZ_classes}

The ten Altland-Zirnbauer symmetry classes are characterised by whether the Hamiltonian satisfies some combination of three types of ``symmetry'': chiral ($\mathcal{C}$), time reversal ($\mathcal{T}$), and particle-hole ($\mathcal{P}$) symmetry. Although these are (anti-)unitary (anti-)symmetries, we will simply refer to them as symmetries. 

Chiral symmetry is characterised by a unitary operator, $\mathcal C = \u{C}$ which anticommutes with the Hamiltonian,
\begin{align}
    \{H, \mathcal C \} = 0.
\end{align}
This constrains all eigenstates to come in pairs with energy $\pm \epsilon$. The chiral symmetry operator $U_C$ can always be chosen to satisfy $U_C^2 = \1$ and thus defines a bipartition of the system labelled by its eigenvalues $\in \{\pm 1\}$. 

Time-reversal and particle hole are both anti-unitary; represented with a combination of a unitary matrix ($\u{A}$) and complex-conjugation ($\mathcal K$),
\begin{equation}
    \mathcal{A} = \u{A} \mathcal K,
\end{equation}
with $\mathcal{A}=\mathcal{T}$ or $\mathcal{P}$. Both of these come in two forms depending on the square of the operator, $\mathcal{A}^2 = \pm 1$, which constrain the unitary part $\u{ A}$ to either be symmetric ($+1$) or antisymmetric ($-1$). 

In the case of time-reversal symmetry, the operator $\mathcal T$ commutes with the Hamiltonian,
\begin{align}
    [ H, \mathcal T] = 0,
\end{align}
whereas a system that is particle-hole symmetric anticommutes with $\mathcal P$,
\begin{align}
    \{ H, \mathcal P\} = 0.
\end{align}
When both time-reversal and particle-hole are present the system also satisfies a chiral symmetry, given by the product of $\mathcal T$ and $\mathcal P$, $\u{C} = \u{P} \u{T}^*$. 

Put together, these symmetries allow for ten Altland-Zirnbauer classes, defined by the presence and sign (i.e.~the value of $\mathcal A^2$) of these three symmetries. They are separated in two “complex” classes (A, AIII), which do not have any antiunitary symmetries, and eight “real” classes, which have at least one antiunitary symmetry, each occupying a row in \cref{tab:tenfold_way}.


\begin{table}[t]
    \centering
        \begin{tabular}{c|ccc|cccccccc}
    \textrm{class} & $\mathcal{C} (= \mathcal {PT})$ & $\mathcal{T}$ & $\mathcal{P}$ & $d=0$ & $1$ & $2$ & $3$ & $4$ & $5$ & $6$ & $7$ \\
    \hline 
    \textrm{A}   &     &      &      & $\mathbb{Z}$   &                & $\mathbb{Z}$   &                & $\mathbb{Z}$   &                & $\mathbb{Z}$   &                \\
    \textrm{AIII}& $1$ &      &      &                & $\mathbb{Z}$   &                & $\mathbb{Z}$   &                & $\mathbb{Z}$   &                & $\mathbb{Z}$   \\
    \hline                               
    \textrm{AI}  &     & $1$  &      & $\mathbb{Z}$   &                &                &                & $2\mathbb{Z}$  &                & $\mathbb{Z}_2$ & $\mathbb{Z}_2$ \\
    \textrm{BDI} & $1$ & $1$  & $1$  & $\mathbb{Z}_2$ & $\mathbb{Z}$   &                &                &                & $2\mathbb{Z}$  &                & $\mathbb{Z}_2$ \\
    \textrm{D}   &     &      & $1$  & $\mathbb{Z}_2$ & $\mathbb{Z}_2$ & $\mathbb{Z}$   &                &                &                & $2\mathbb{Z}$  &                \\
    \textrm{DIII}& $1$ & $-1$ & $1$  &                & $\mathbb{Z}_2$ & $\mathbb{Z}_2$ & $\mathbb{Z}$   &                &                &                & $2\mathbb{Z}$  \\
    \textrm{AII} &     & $-1$ &      & $2\mathbb{Z}$  &                & $\mathbb{Z}_2$ & $\mathbb{Z}_2$ & $\mathbb{Z}$   &                &                &                \\
    \textrm{CII} & $1$ & $-1$ & $-1$ &                & $2\mathbb{Z}$  &                & $\mathbb{Z}_2$ & $\mathbb{Z}_2$ & $\mathbb{Z}$   &                &                \\
    \textrm{C}   &     &      & $-1$ &                &                & $2\mathbb{Z}$  &                & $\mathbb{Z}_2$ & $\mathbb{Z}_2$ & $\mathbb{Z}$   &                \\
    \textrm{CI}  & $1$ & $1$  & $-1$ &                &                &                & $2\mathbb{Z}$  &                & $\mathbb{Z}_2$ & $\mathbb{Z}_2$ & $\mathbb{Z}$   \\
    \hline
    \end{tabular}
    \caption{Tenfold-way classification of strong topological phases by symmetry class and dimension.}
    \label{tab:tenfold_way}
\end{table}

\subsection{Tenfold-way invariants in 0D}
\label{sec:invariants_in_0d}

Of the ten AZ symmetry classes, five can host topological invariants for a given dimensionality. In zero dimensions, these phases can be characterised by two types of invariant: the signature (in classes A, AI, and AII) and sign of the Pfaffian (in classes D, BDI).

In classes A, AI, AII, topological distinct phases are simply ones with different numbers of occupied states---i.e.~eigenstates with negative energy. The invariant is the half signature of the Hamiltonian, defined as half the difference between the number of positive and negative eigenvalues of $H$,
\begin{align}
    \frac 12 \sig(H) \nonumber = \frac 12 (n_+ - n_-) \in \mathbb{Z}. 
\end{align}
$H$ is Hermitian, so all eigenvalues are real. Thus, the only way to continuously transform $H$ into a Hamiltonian $H'$ with a different signature is if an eigenvalue crosses zero at some point in the deformation. Note that in class AII time-reversal symmetry with $\mathcal T^2 = -1$ leads to Kramers' degeneracy thus forcing the invariant to be even since all eigenvalues are doubly degenerate, yielding a $2\mathbb{Z}$, rather than $\mathbb{Z}$ classification. 

In classes D and BDI the Hamiltonian is constrained by particle-hole to have eigenvalues that come with opposite energy pairs---thereby forcing the signature to vanish. These classes, however, have particle-hole symmetry (with $\mathcal P^2 = +1$) and can thus be characterised by a Pfaffian. The antisymmetry of $\u{ P}$ ensures that the product $H \u{P}$ is also anti-symmetric,
\begin{align}
\begin{aligned}
    H \u{P} &= - \u{P} H^* \\
    &= - (H \u{P})^T,
\end{aligned}
\end{align}
where we used $\u{P}=\u{P}^T$. The sign of the Pfaffian,
\begin{align} \label{eqn:h_up}
    \sgn \pf(i H \u{P}) \in \mathbb{Z}_2,
\end{align}
is therefore a topological invariant\footnote{Class DIII also has $\mathcal P^2 = +1$, but the presence of Kramer's degeneracy fixes the sign of the Pfaffian to always be the same.}.
Note that in these classes we may rotate to the Majorana basis, \textit{i.e.} when $\u{P} = \1$, the Hamiltonian is anti-symmetric (and purely imaginary) and the invariant is simply $\sgn \pf(iH)$. Typically it is more convenient to use \cref{eqn:h_up}.

Finally, in class BDI we can take advantage of the additional chiral symmetry to express this invariant as a determinant rather than the (numerically more costly) Pfaffian. By rotating to a basis where $\u{C} = \sigma_3$ and $\u{P}=\1$ the Hamiltonian is an antisymmetric block off-diagonal matrix,
\begin{align}
    H = \begin{pmatrix}
        & i B \\
        -iB^T &
    \end{pmatrix},
\end{align}
where $B$ is a real matrix. We make use of the following identity, 
\begin{align}\begin{split}
    \sgn \pf(i H) &= (-1)^{{N(N+1)}/2} \sgn \det(B) \in \mathbb Z_2.
\end{split}
\end{align}
Usually this dimensionality-dependent factor of $(-1)$ is neglected since the simple determinant of $B$ still gives a valid $\mathbb Z_2$ invariant. 

\subsection{Additional unitary symmetries}

We next consider imposing a group of \emph{unitary} symmetries $G \subseteq U(N)$, in addition to AZ symmetries. 
The Hamiltonian $H$ can be block diagonalized \cite{heinzner2005symmetry, morimoto2013} into sectors $H_\alpha$ labelled
\footnote{Note that strictly speaking if $\alpha$ is a $d_\alpha$-dimensional irrep of $G$ then $H_\alpha$ is $d_\alpha$ fold degenerate and the decompostion looks like $$H = \bigoplus_\alpha \1_{d_\alpha} \otimes H_\alpha$$.  }
by the irreducible representations (irreps) $\alpha$ of $G$,
\begin{equation}
    H = \bigoplus_\alpha \1_{d_\alpha} \otimes H_\alpha.
\end{equation}
The AZ symmetries either act within a sector, in which case $H_\alpha$ inherits that AZ symmetry, or relates a pair of sectors $\alpha, \beta$, in which case $H_\alpha$ and $H_{\beta}$ do not inherit the AZ symmetry.
The symmetry class of $H$ is thus specified by the algebraic structure of the imposed symmetries: the unitary symmetry group $G$, the AZ symmetries, and the action of the AZ symmetries on $G$.

We can then classify each block $H_\alpha$ individually according to the AZ symmetries that it inherits. The topological invariants are then signatures or Pfaffians in each block individually.

As an example consider $H$ with chiral symmetry $\mathcal{C}$ (class AIII) and inversion $\mathcal{I}$. $H$ can be be split into $\pm$ parity sectors based on the \textit{inversion} eigenvalues,
\begin{align}
    H &= 
    \begin{pmatrix}
    H_+ & \\
    & H_-
    \end{pmatrix}.
\end{align}
If $[\mathcal{C},\mathcal{I}]=0$ then $\mathcal{C}$ acts within a parity sector so $H_\pm$ are class AIII Hamiltonians and thus have a trivial classification. 
If instead $\{\mathcal{C},\mathcal{I}\}=0$ then $\mathcal{C}$ maps between parity sectors so $H_\pm$ are class A Hamiltonians with opposite spectra, and with an integer classification
\begin{equation}
    \frac12 \sig{H_+} = - \frac12 \sig{H_-} \in \mathbb{Z}.
\end{equation}
The two different algebras, in which $\mathcal{C}$ and $\mathcal{I}$ either commute or anti-commute, correspond to distinct symmetry classes, sometimes denoted as $\textrm{AIII}^{\mathcal{I}_+}$ and $\textrm{AIII}^{\mathcal{I}_-}$, respectively\cite{geier2018second,trifunovic2019higher}.

\subsection{Additional anti-unitaries}

We can also treat more general anti-unitaries $\mathcal{A}$ (\emph{e.g.} $C_4\mathcal{T}$ or $C_6\mathcal{P}$). Much like time-reversal and particle-hole symmetries, these anti-unitaries come with two variants
\begin{equation}
    \mathcal{A}^{n} = \pm \1.
\end{equation}
Here $n$ is the smallest integer where the power is proportional to the identity, for example $n = 4$ for $C_4\mathcal{T}$.
As was the case with unitary symmetries we start by block diagonalizing $H$, this time using the ``unitary halving subgroup''\cite{wigner1959group,bradley2009mathematical,bradley1968magnetic,zhang2015} of the anti-unitary, \textit{i.e.} we block diagonalize with respect to the unitary symmetry $\mathcal{A}^2$. If $\mathcal{A}$ is an (anti-)symmetry of $H$ then $\mathcal{A}^2$ is a symmetry of $H$ and we can block-diagonalize into sectors $H_\alpha$ labelled by the irreps $\alpha$ of $\mathcal{A}^2$. The antiunitary $\mathcal{A}$ then acts like an order-two symmetry that either maps between blocks or acts within a block.

As examples, we consider class A Hamiltonians with $C_4\mathcal{T}$. The two cases of $(C_4\mathcal{T})^4=\pm1$ define distinct symmetry classes which we treat separately.

\emph{Example $(C_4\mathcal{T})^4=+1$}: 
We can block-diagonalize $H$ into blocks labelled by parities $\pm1$ under $C_2 = (C_4\mathcal{T})^2$. Within a given block we have $(C_4\mathcal{T})^2=\pm1$ 
\begin{equation}
    H =  
    \begin{pmatrix}
    H_+ & \\
    & H_- \\
    \end{pmatrix}, \quad
    C_4\mathcal{T} =  
    \begin{pmatrix}
    U_+ & \\
    & U_- \\
    \end{pmatrix}K,
\end{equation}
where $U_\pm = \pm U_\pm^T$. We thus have $H_+$ that is in class AI and $H_-$ in class AII. These two blocks carry a $\mathbb{Z}$ and a $2\mathbb{Z}$ invariant
\begin{equation}
    (\sig H_+, \sig H_-) \in \mathbb{Z}\times2\mathbb{Z}.
\end{equation}

\emph{Example $(C_4\mathcal{T})^4=-1$}: 
We again block-diagonalize $H$ but this time with blocks labelled by $C_2$ eigenvalues $\pm i$,
\begin{equation}
    H =  
    \begin{pmatrix}
    H_{+i} & \\
    & H_{-i} \\
    \end{pmatrix}, \quad
    C_4\mathcal{T} =  
    \begin{pmatrix}
    0 & U \\
    -i U^T & 0 \\
    \end{pmatrix}K.
\end{equation}
where the form of $C_4\mathcal{T}$ is set by working in the basis where $(C_4\mathcal{T})^2=i\sigma_3$. In this case $C_4\mathcal{T}$ maps between blocks
\begin{align}
    \begin{pmatrix}
    H_{+i} & \\
    & H_{-i} \\
    \end{pmatrix} 
    &=   
    \begin{pmatrix}
    U H_{-i}^* U^\dag & \\
    & U^T H_{+i}^* U^*\\
    \end{pmatrix},
\end{align}
and we're left with two Class A Hamiltonians which are time-reversal copies of one another. The Hamiltonian is the characterise my a single integer
\begin{equation}
    \sig H_{+i} = \sig H_{-i} \in \mathbb{Z}.
\end{equation}

\section{The localizer: beyond zero dimensions}\label{sec:dim_recg_and_cliff}

Having reviewed the topological classification of Hamiltonians in zero dimensions, we turn to the classification of Hamiltonians in an $d$-dimensional system, where the notion of topological equivalence changes. Now, we not only have a matrix $H$, but also a notion of position in physical space associated with it, where each dimension adds a corresponding position operator $R_j$. Along with these position operators comes the constraint of \textit{locality}: a physical Hamiltonian should not be able to hop particles an arbitrary distances across space. Thus, our topological criteria posed at the start of \cref{sec:zero_dimensions} must also change. Two Hamiltonians, $H$ and $H'$ are considered equivalent if they can be connected without closing the gap \textit{or} breaking locality at any point in the deformation. In an $d$-dimensional system, we thus have $d+1$ constraints to impose, a gap constraint and a locality constraint for every dimension.

It is the addition of these extra locality constraints that makes topological invariants in higher-dimensional systems much more complex than the zero-dimensional topology reviewed in the previous section. It is not at all obvious how one should design an invariant that can detect a breaking of locality on an equal footing to a breaking of gapped-ness---the $d$-dimensional question is simply more complex.\footnote{From a momentum-space perspective, this locality constraint is always implied by the assumption that the band be continuous---i.e.~that the energy $E(\textbf k)$ is a continuous function of $\textbf k$.}

The object central to our discussion will be the spectral localizer, constructed from the Hamiltonian and position operators,
\begin{equation} \label{eqn:TheLocalser}
    \mathcal{L} = H \otimes \Gamma_{d+1} + \kappa \sum_{i=1}^d R_i \otimes \Gamma_i,
\end{equation}
where $\Gamma_i$ are anti-commuting matrices that form a Clifford algebra $\{\Gamma_i, \Gamma_j\} = 2\delta_{ij}$, and $\kappa$ is a scaling parameter that ensures $H$ and $R_i$ have compatible units\footnote{Note that the localizer is often introduced with position and energy shifts $\mathcal{L} = (E_F \1 -  H) \otimes \Gamma_{d+1} + \kappa \sum_{i=1}^d (r_i\1 - R_i) \otimes \Gamma_i$. We will take these to be zero throughout, $E_F=0,\ \vec{r}=0$, thus assuming the Fermi-energy to be at zero-energy and the position operators to be centred about the origin.}. Going forward we will often drop the tensor products ($\otimes$) to avoid notational clutter.

In this section, we shall show how $\mathcal L$ is an an object whose gap encodes both the gapped and local nature of $H$ --- \emph{i.e.} the localizer becomes gapless if $H$ is \textit{either} gapless \textit{or} non-local. This allows us to use the localizer $\mathcal L$ to effect a dimensional reduction of our problem: rather than asking whether the Hamiltonian $H$ can be transformed into $H'$ preserving gappedness and locality, we ask whether the corresponding $\mathcal L$ and $\mathcal L'$ can be transformed into one another preserving gapped-ness only---effectively reducing the problem of $d$-dimensional topology into the zero-dimensional problem that we already solved in \cref{sec:zero_dimensions}.

This section has three parts. 
In \cref{sec:encoding_locality} we provide a definition of locality, and show how show that the spectrum of $\mathcal{L}$ encodes information about both the gap of the Hamiltonian and its locality. We make no pretence at rigour, leaving more rigorous treatments~\cite{loring_algebras_2015, cerjan_classifying_2024} to the existing literature~\cite{loring_guide_2019,loring_spectral_2019, loring_spectral_2020, schulz-baldes_spectral_2021,doll_skew_2021, schulz_invariants_2022, cerjan_operator-based_2022,cerjan_local_2022, schulz_spectral_2023, cerjan_spectral_2023, franca_topological_2024, cerjan_crystal_2024, doll_local_2024, schulz_topological_2024, stoiber_spectral_2024, lee_classification_2025, wong_efficient_2026}, and instead focus on providing intuition. In order to construct the relevant invariant for a 0D system, we must know its symmetry class. Thus, in \cref{sec:loc_tenfold}, we discuss how the symmetry class of the localizer can be determined from the symmetry class of the starting Hamiltonian by Bott periodicity. 
To that end, we introduce a recursive dimensional reduction scheme, starting with $H$ and terminating with $\mathcal{L}$, that constructs a series of localizers that characterise all weak and strong invariants in a given symmetry class. 
This gives a unifying structure to the localizer index across all AZ symmetry classes, and sets the foundation for the inclusion of real-space symmetries in the next section. 
Finally, having shown that this localizer methodology reproduces the established classification for the tenfold way, in \cref{sec:dirac_loc} we apply the spectral localizer to the canonical Dirac Hamiltonians describing the low-energy spectrum around a topological phase transition. 
We show that a change in $H$'s invariant is accompanied by an equivalent change in $\mathcal{L}$'s invariant, verifying that the that the invariants constructed with  $\mathcal{L}$ are precisely aligned with the tenfold-way classification of $H$. 

Together, these highlight how the spectral localizer is an excellent tool for both \emph{classification} and \emph{characterization}. Providing a simple, general machinery to obtain both the classifying group and the topological invariant of a symmetry class.

\subsection{Encoding locality} \label{sec:encoding_locality}
In $d$-dimensions, the Hamiltonian can be represented by separating the real-space and on-site components,
\begin{align}
    H = \sum_{\alpha \beta} \ketbra{\bm r_\alpha }{\bm r_\beta} \otimes h_{\alpha \beta},
\end{align}
where $\ket{\bm r_\alpha}$ are a set of position eigenstates, and the hopping element $h_{\alpha \beta}$ acts on the on-site (or orbital) degrees of freedom. There are various definitions one could construct for locality~\cite{lieb_finite_1972,hastings_locality_2010}, here we use a reasonably weak one.
For each spatial dimension ($X, Y, Z ... \in \vec{R}$ ), all entries in the hopping element $h_{\alpha \beta}$ decay at least as fast as a power law in their distance,
\begin{align}
    \max (h_{\alpha \beta}) \leq \frac \mu {|x_\alpha- x_\beta|}, 
\end{align}
where $\mu$ is a finite constant that we choose, and the constraint also applies for $y_\alpha$, $z_\alpha$ and so on. For a system in infinite boundary conditions this definition is sufficiently strong for any finite $\mu$. However, when working in a finite system of size $L$, the simplest requirement is to ask that $\mu \ll L \epsilon_{\textup{max}}$, the size of the system multiplied by the bandwidth, see Refs.~\cite{loring_algebras_2015, cerjan_classifying_2024} for a more detailed discussion.
As we show in \cref{apx:locality}, this constraint is equivalent to requiring that the velocity operator, given by the commutator between $H$ and $X$, has its spectrum bounded by a small parameter $\mu$---also referred to as \textit{almost commuting}~\cite{hastings_almost_2010}, 
\begin{align} \label{eqn:HX_comm}
    \parallel -i [H, X] \parallel\ \leq \mu,
\end{align}
where $\parallel * \parallel$ denotes the spectral radius (or 2-norm) given by the extremal absolute value of any eigenvalue. 

On the face of it, these two constraints---locality and gapped-ness---appear completely inequivalent, and our aim is to find a quantity that can simultaneously detect a violation in either. 
To see heuristically how $\mathcal L$ encodes encodes both locality and gapped-ness consider the localizer squared,
\begin{align}
    \mathcal L^2 =  [H^2 + (\kappa R_j )^2]\otimes \1 + \kappa 
    [H,  R_j]
    \otimes
    \Gamma_{d+1} \Gamma_{j},
\end{align}
where we sum over the $j$ index. Given the Hamiltonian is assumed to have a gap of $\Delta$, we can bound the spectrum of the first term from below by $\Delta^2$. Similarly, using \cref{eqn:HX_comm}, we can bound the spectrum of each of the $d$ commutator terms between $\pm \kappa\mu$. Thus, Weyl's inequality ensures that the gap of $\mathcal L^2$ satisfies
\begin{align}
    \textrm{gap}(\mathcal L^2) \geq \Delta^2 - d \kappa \mu.
\end{align}
The only way for the localizer to become gapless is \textit{either} to close the gap of the Hamiltonian, reducing $\Delta$, or to break locality, increasing $\mu$ (one could of course perform some combination of both as well). Thus, we see that the spectral gap of the localizer encodes information about both \textit{both} the gapped-ness and the locality of $H$. When we try to build a smooth path $H(\lambda)$ between two Hamiltonians, $H$ and $H'$, the topological equivalence can be detected by studying the corresponding spectral localizer $\mathcal L(\lambda)$. As long as $\mathcal L(\lambda)$ remains gapped throughout the transformation, we have the guarantee that we neither broke gappedness, nor locality of $H(\lambda)$ and the two Hamiltonians $H$ and $H'$ are topologically equivalent.

The path to an invariant is now clear. We start with a $d$-dimensional problem: can two Hamiltonians be connected without closing a gap or breaking locality. We then reduce it to an equivalent 0-dimensional question: Can two \textit{localizers} be connected without \textit{closing the gap only}. This can be studied by treating the localizer itself as though it were a zero-dimensional Hamiltonian, and constructing the invariant using the methods reviewed in \cref{sec:invariants_in_0d}. 

\subsection{Dimensional reduction: the Bott clock} \label{sec:loc_tenfold}

Having cast the localizer $\mathcal{L}$ as a `zero-dimensional' proxy for the topology of $H$, we can apply the classification methodology laid out in \cref{sec:zero_dimensions}. We must, however, first understand which symmetry class the localizer operator itself falls into. Here we consider the ten AZ symmetry classes, showing how the symmetries of the localizer follow from those of the Hamiltonian. We leave the treatment of crystalline symmetries to the next section.

The symmetries of the localizer $\mathcal L$ depend both on the symmetries of the original Hamiltonian $H$, as well as the number of position operators introduced to produce $\mathcal L$. 
Here we show how, for every position operator included in the localizer, our symmetry class takes a single step along the Bott clock.
That is, the localizer's class follows
\begin{align} \label{eqn:complex_order}
    \textup{A} \rightarrow \textup{AIII} \rightarrow \textup{A} \rightarrow \textup{AIII} \rightarrow  \cdots
\end{align}
for the complex classes and 
\begin{align}\label{eqn:real_order}
    \textup{CI} \rightarrow \textup{C}\rightarrow \textup{CII}\rightarrow \textup{AII} \rightarrow \cdots \rightarrow \textup{AI} \rightarrow\textup{CI}\rightarrow  \cdots
\end{align}
for the eight real classes---effectively following the `upper left diagonal' in \cref{tab:tenfold_way}.
This is closely related to how Bott periodicity arises in the classification of topological defects by Teo and Kane~\cite{teo2010topological}. 


We show this by first considering the symmetry class of a localizer constructed from a Hamiltonian $H$ and a \textit{single} position operator, $X$---effectively dimensionally reducing by one. 
Unlike the existing literature~\cite{loring_algebras_2015, cerjan_classifying_2024}, we propose a different localizer step when the Hamiltonian $H$ has chiral symmetry than when it does not. 
In \cref{sec:multiple_loc}, we extend this procedure to show how one can reduce an arbitrary number of dimensions. This is done by iteratively applying the single localizer step, at each point treating the localizer obtained after the the last step as though it were the Hamiltonian to be used for the next one. This yields a spectral localizer for any AZ symmetry class in any dimension.

\subsubsection{Reducing a single dimension}\label{sec:one_step}


\textit{Without chiral symmetry---} When the Hamiltonian does \emph{not} have chiral symmetry the localizer can be constructed as
\begin{align} \label{eqn:loc_without_c}
    \mathcal L = 
    H \sigma_2 +
    \kappa X \sigma_1.
\end{align}
Where the Pauli matrices $\sigma_i$ are the Clifford matrices that act on an auxiliary two-dimensional space introduced in the localizer construction. 
First, notice that the localizer has chiral symmetry by construction; it anticommutes with the unused Pauli matrix $\mathcal C = \sigma_3$. 

If the Hamiltonian has no other symmetries we find that a class A Hamiltonian has a class AIII localizer, 
\begin{align}
\textup{A} \rightarrow \textup{AIII}. 
\end{align}

If the Hamiltonian has time-reversal symmetry, with an anti-unitary operator $\mathcal T =  \u T K$ that squares to $\eta_{\mathcal T} \in \{\pm 1\}$, then we can construct both a $\widetilde {\mathcal{T}}$ and a $\widetilde {\mathcal{P}}$ operator which are symmetries of the localizer, given by
\begin{align}
    \widetilde {\mathcal T} &= \u T \sigma_1 K, \quad \textup{with } \widetilde {\mathcal T}^2 = +\eta_{\mathcal T},\\
    \widetilde {\mathcal P} &= \u T \sigma_2 K,  \quad \textup{with } \widetilde {\mathcal P}^2 = -\eta_{\mathcal T}.
\end{align}
Thus, labelling the symmetry as $(\eta_{\mathcal T},\eta_{\mathcal P})$, we find that the symmetry class changes as
\begin{align}
    (+1,0) &\rightarrow (+1,-1), \quad \textup{AI} \rightarrow \textup{CI}, \\
    (-1,0) &\rightarrow (-1,+1), \quad \textup{AII} \rightarrow \textup{DIII}.
\end{align}

If the Hamiltonian instead has a particle-hole symmetry, with an anti-unitary operator $\mathcal P =  \u P K$ that squares to $\eta_{\mathcal P} \in \{\pm 1\}$, then again we construct both a $\widetilde {\mathcal{T}}$ and $\widetilde {\mathcal{P}}$ symmetry, which now have the form
\begin{align}
    \widetilde {\mathcal T} &= \u P \sigma_0 K, \quad \textup{with } \widetilde {\mathcal T}^2 = +\eta_{\mathcal P},\\
    \widetilde {\mathcal P} &= \u P \sigma_3 K, \quad \textup{with } \widetilde {\mathcal P}^2 = +\eta_{\mathcal P},
\end{align}
which lead to a symmetry change of 
\begin{align}
    (0, +1) &\rightarrow (+1,+1), \quad  \textup{D} \rightarrow \textup{BDI}, \\
    (0, -1) &\rightarrow (-1,-1), \quad  \textup{C} \rightarrow \textup{CII}.
\end{align}

\textit{With chiral symmetry---} When systems have chiral symmetry the Hamiltonian already has a Clifford structure which must be taken into account, where the chiral operator acts as an unused Clifford matrix. In the basis where chiral symmetry is $\mathcal{C} = \sigma_3$ the Hamiltonian takes the form,
\begin{align}
    H =  H_+ \otimes \sigma_1 +  H_{-} \otimes \sigma_2.
\end{align}
Where we have temporarily restored the tensor products $\otimes$ for clarity. 
The spectral localizer is constructed by using chiral symmetry in place of the missing Clifford operator
\footnote{
    Note that we can arrive at this form of the localizer even if we do not exploit $H$'s Clifford structure and instead work with $\mathcal{L}=H\otimes\tau_3+X\otimes\tau_1$. This localizer has chiral symmetry $\tau_2$ and a unitary symmetry $U=\mathcal{C}\otimes\tau_1$. In the $U=+1$ sector, the $\tau_1$ eigenvalue is locked to the $\mathcal{C}$ eigenvalue, so that $X\otimes\tau_1 \mapsto \mathcal{C}X$, while $H\otimes\tau_3\mapsto H$. This symmetry-reduced localizer\cite{cerjan_classifying_2024} $\mathcal{L}_+ = H + \mathcal{C}X$ is an alternative route to \cref{eqn:loc_with_c}.
},
\begin{align}\label{eqn:loc_with_c}\begin{aligned}
    \mathcal{L} 
    &= H + \mathcal{C}X,\\ 
    & = H_{+}\otimes \sigma_1 + H_{-}\otimes \sigma_2 +  \sigma_3 X.
\end{aligned}
\end{align}
Note that $\mathcal{C}X$ is the product of the chiral and position operators, not a tensor product. 
This localizer breaks the chiral symmetry that we started with. 
If the Hamiltonian has no other symmetries then we find a symmetry class change of
\begin{align}
    \textup{AIII} 
    \rightarrow 
    \textup{A}.
\end{align}

We next consider the case where $H$ further has both time-reversal and particle-hole symmetries, each squaring to $\eta_{\mathcal T}$ and $\eta_{\mathcal P}$ respectively. 
Since chiral symmetry is broken in $\mathcal L$ by the $\mathcal{C}X$ term, it must also break at least one of $\mathcal T$ or $\mathcal P$. 
To see which is broken, note that $X$ commutes with all three of $\mathcal{T},\mathcal{P},\mathcal{C}$, and that $\mathcal{C}$ has the same commutation relation with both $\mathcal{T}$ and $\mathcal{P}$,
\begin{align}\label{eqn:PT_comm}
    \mathcal T \mathcal C &= \eta_{\mathcal T} \eta_{\mathcal P} \mathcal C \mathcal T \nonumber\\
    \mathcal P \mathcal C &= \eta_{\mathcal T} \eta_{\mathcal P} \mathcal C \mathcal P. 
\end{align} 
It follows that when $\eta_{\mathcal T} \eta_{\mathcal P} = +1$ the localizer preserves $\mathcal{T}$ but breaks $\mathcal{P}$, inducing the following symmetry changes,
\begin{align}
    (+1, +1) &\rightarrow (+1,0),\quad  \textup{BDI} \rightarrow \textup{AI}, \\
    (-1, -1) &\rightarrow (-1,0),\quad  \textup{CII} \rightarrow \textup{AII}.
\end{align}
Conversely, when $\eta_{\mathcal T} \eta_{\mathcal P} = -1$ the localizer preserves $\mathcal{P}$ but breaks $\mathcal{T}$,
\begin{align}
    (+1, -1) &\rightarrow (0,-1),\quad  \textup{CI} \rightarrow \textup{C}, \\
    (-1, +1) &\rightarrow (0,+1),\quad  \textup{DIII} \rightarrow \textup{D}.
\end{align}
Altogether these results reproduce the Bott clock ordering given in \cref{eqn:complex_order,eqn:real_order}. 


\subsubsection{Strong and weak phases} \label{sec:multiple_loc}

We now turn to how to add multiple position operators. 
Given a $d$-dimensional Hamiltonian in a given symmetry class and the corresponding set of $d$ position operators, we can iteratively apply the two steps provided above, each time feeding the localizer produced in the previous step into the next step as though it were a new Hamiltonian. This enables the construction of a localizer in the $0d$ symmetry class with the same classification as the $d$-dimensional Hamiltonian's. 


We illustrate this process with an example, that of a $3d$ Hamiltonian in class AII (\emph{i.e.} $\mathcal{T}^2=-1$), which has a $\mathbb Z_2$ invariant.

The first step is to include a single position operator, $X$, following \cref{eqn:loc_without_c},
\begin{align}
    \mathcal L_X = H \sigma_2 + \kappa  X\sigma_1.
\end{align}
This sits in class DIII. 
Having only added a single position operator, $\mathcal{L}_X$ only encodes locality in the x-direction. 
It therefore probes the topology of $H$ as if it were a $1d$-wire, with a huge unit-cell in the $y, z$ directions (in this case there are no topological phases as class DIII is trivial in $0d$\footnote{Equally one can note that class AII is trivial in $1d$.}). 
We could equally well have constructed $\mathcal{L}_Y, \mathcal{L}_Z$. 

As $\mathcal{L}_X$ has chiral symmetry, we use \cref{eqn:loc_with_c} to produce the second localizer,
\begin{align}
    \mathcal L_{XY} = H \sigma_2 + \kappa  X \sigma_1 + \kappa  Y \sigma_3,
\end{align}
which is in class D. 
With two position operators, $\mathcal{L}_{XY}$ probes the topology of $H$ as if it were a $2d$-system: calculating the quantum-spin Hall invariant of the $3d$ system as if it were a thick $2d$ one.

Finally, since $\mathcal L_{XY}$ doesn't have chiral symmetry, the third localizer follows from \cref{eqn:loc_without_c}, 
\begin{align}\label{eqn:unwieldy}
    \mathcal L_{XYZ} 
    &= H \sigma_2 \tau_3 + \kappa X \sigma_1 \tau_3 + \kappa Y \sigma_3 \tau_3 + \kappa Z \sigma_0 \tau_1,
\end{align}
which is in class BDI. This localizer's invariant (the Pfaffian) is the one that probes the $3d$ system's $\mathbb{Z}_2$ topology, diagnosing whether the system is a trivial insulator or a $3d$ TI.

This localizer is of the form of \cref{eqn:TheLocalser},
\begin{align}\label{eqn:unwieldy}
    \mathcal L_{XYZ} 
    &= H \Gamma_4 + \kappa \sum_i R_i \Gamma_i,
\end{align}
Where the tensor products of Pauli matrices (i.e.~$\sigma_i \tau_j$) form four elements of a five-element representation of the Clifford algebra, which we relabel $\Gamma_i$. 
Note that this feature, that successive applications of a single localizer step automatically generates higher-order representations of the Clifford algebra, is shown in \cref{apx:pauli_building}. 

This iterative procedure generates a sequence of localizers, each treated as a $0d$-object in a different symmetry class, encoding locality in the directions whose position operators were introduced. 
In each case, we construct the $0d$ invariant in that symmetry class as laid out in \cref{sec:invariants_in_0d}, 
\begin{alignat}{2}
    H
    &\in{}& \makebox[2em][c]{\textup{AII}}
    &\implies \frac{1}{2}\sig H \in 2\mathbb Z, \\
    \mathcal{L}_X, \mathcal{L}_Y, \mathcal{L}_Z
    &\in{}& \makebox[2em][c]{\textup{DIII}}
    &\implies \textup{no invariant}, \\
    \mathcal{L}_{XY}, \mathcal{L}_{YZ}, \mathcal{L}_{ZX}
    &\in{}& \makebox[2em][c]{\textup{D}}
    &\implies \sgn \pf \mathcal{L}_{**} \in \mathbb Z_2, \\
    \mathcal{L}_{XYZ}
    &\in{}& \makebox[2em][c]{\textup{BDI}}
    &\implies \sgn \pf \mathcal{L}_{XYZ} \in \mathbb Z_2.
\end{alignat}

Of these, $\mathcal{L}_{XYZ}$ diagnoses \emph{strong} phases. These are classified in \cref{tab:tenfold_way} and are the phases one normally associates with topology in the tenfold-way symmetry classes. These are phases that do \emph{not} need translation symmetry for their protection. From a classification perspective, this corresponds to the K-group on the sphere\cite{kitaev2009periodic,ryu_topological_2010}. 

The intermediate localizers diagnose phases as if the system were lower dimensional. 
If our system further has translation symmetry in a direction whose position operator was not included, these intermediate localizers can be used to probe \emph{weak topological phases}. 
These are phases that require translation symmetry for their protection and their classification requires the K-group on the torus~\cite{kitaev2009periodic,ryu_topological_2010}. 
Their weak invariants can be obtained by evaluating lower-dimensional topological invariants on lower-dimensional cuts of the Brillouin zone~\cite{fu2007three,moore2007topological,fu2007inversion}.
For example, consider $\mathcal{L}_{XY}$ for a system with translation symmetry along $z$. 
To incorporate translation symmetry along $z$, we consider $H(k_z)$, obtained from a single unit cell in $z$ with twisted boundary conditions, leaving a two-dimensional system in $x$ and $y$ with no assumption of translation symmetry in those directions (\emph{i.e.}, a partial Fourier transform only in $z$). 
We can then probe weak invariants from 
\begin{equation}
    \mathcal{L}_{XY}(k_z) = H(k_z)\sigma_2 + \kappa(X \sigma_1 + Y \sigma_3).
\end{equation}
This probes whether the model is deformable to a stack of quantum-spin Halls~\cite{fu2007three,moore2007topological}.

Together, this sequence of localizers is able to probe all the strong and weak phases of a Hamiltonian $H$ in an AZ symmetry class.

\subsection{Dirac Localizer} \label{sec:dirac_loc}

As a final consideration before we move on to incorporating spatial symmetries we analyze the localizer for a Dirac Hamiltonian,  aimed at readers more familiar with continuum theories. This is an example of a simple setting where the spectrum of the localizer is analytically tractable, and serves as a demonstration that a change in $H$'s invariant is accompanied by a change in $\mathcal{L}$'s invariant---i.e.~the localizer is probing exactly the same topological phases as the established tenfold way classification. To our knowledge, this sort of result was first considered in Ref.~\cite{schulzbaldes2023spectral}.

In the vicinity of a topological phase transition, the bulk Hamiltonian can be described by
\begin{equation}
    H \simeq m\gamma_{0} + v \vec{k}\cdot\vec{\gamma},
\end{equation}
where $\gamma_i$ form a Clifford algebra. 
When $H$ is in a symmetry class that hosts topological phases, $m\gamma_{0}$ is a unique, symmetry-allowed, anti-commuting mass term that controls which phase the system is in~\cite{kitaev2009periodic}. 
The localizer takes the form
\begin{align}
    \mathcal{L} =  m\gamma_{0}\Gamma_{0} + v\vec{k}\cdot\vec{\gamma}\Gamma_{0} + \kappa \vec{R}\cdot\vec{\Gamma}.
\end{align}
In the basis where $\gamma_{0}\Gamma_{0}=\sigma_3$ the localizer can be written as 
\begin{equation}
    \mathcal{L} = 
    \begin{pmatrix}
    m & D \\
    D^\dag & -m
    \end{pmatrix},
\end{equation}
Most states in $\mathcal{L}$'s spectrum come in $\pm$ pairs
\begin{equation}
    \lambda_i = \pm \sqrt{m^2 + d_i^2}
\end{equation}
where $d_i$ are singular values of $D$ (\emph{i.e.} $d_i^2$ are eigenvalues of $D^\dag D$). 
States in the kernel of $D$ or $D^\dag$ (\emph{i.e.} $d_i=0$) do not, however, come in $\pm$ pairs
\begin{equation}
    \lambda =
    \begin{cases}
    -m, & \text{for } \ket{u} \in \ker D, \\
    +m, & \text{for } \ket{v} \in \ker D^\dagger.
    \end{cases}
\end{equation}
with sublattice polarized eigenvectors, $\ket{-m} = (0, \ket{u})^T$ where $D \ket{u} = 0$ or $\ket{+m} = (\ket{v}, 0)^T$ where $D^\dag \ket{v} = 0$. 
It is states that do not have an opposite eigenvalue partner that change the signature of $\mathcal{L}$ upon changing the sign of $m$. 
Changing the sign of $m$, and thus pushing $H$ through a topological transitions, results in an invariant change
\begin{equation}
    \Delta \nu = \frac12 \Delta \sig(\mathcal{L}) = \textrm{ind}(D),
\end{equation}
where $\textrm{ind}(D) = \dim \ker D - \dim \ker D^\dag$ is the topological index.

As an example, we apply this  to the Dirac model of an SHH chain~\cite{su_solitons_1979,Jackiw1976} near the transition $H\simeq m\sigma_3 + v k \sigma_1$ between invariants $\nu=1$ and $\nu=0$. 
The localizer takes the form
\begin{equation}
    \mathcal{L} = m\sigma_3 + v k \sigma_1 + \kappa x \sigma_2.
\end{equation}
The localizer invariant is $\frac12 \sig(\mathcal{L})$. 
Defining the harmonic oscillator annihilation operator $a = (x/l + l \partial_x)/\sqrt{2}$ (where $l=\sqrt{v/\kappa}$, $\omega = \sqrt{2v\kappa}$) we can write
\begin{equation}
    \mathcal{L} = 
    \begin{pmatrix}
    m & -i \omega a\\
    i \omega a^\dag & -m
    \end{pmatrix}.
\end{equation}
The kernel of $a$ is the ground state of the harmonic oscillator $\psi_0(x)\propto e^{-x^2/(2l^2)}$, while $\ker a^\dag=0$. This gives $\textrm{ind}(a) = +1$. Changing from positive to negative $m$ thus results in
\begin{align}
    \Delta \nu 
    &= \frac12 \Delta \sig(\mathcal{L}) = \textrm{ind}(a) = +1. \nonumber
\end{align}
We thus find that the localizer invariant follows that of the Hamiltonian. 

As a second example, we apply this view to the Dirac model of a Chern insulator near the transition $H\simeq m\gamma_0 + v k_x \gamma_1 + v k_y \gamma_2$ between invariants $\nu=1$ and $\nu=0$.
We work in the basis where the localizer takes the form
\begin{align}
    \mathcal{L} =
    m \sigma_3\tau_0
    &+ \kappa x \sigma_1\tau_0 + \kappa y \sigma_2\tau_2 \nonumber \\
    &- v k_x \sigma_2\tau_3 - v k_y \sigma_2\tau_1.
\end{align}
The localizer invariant is $\frac12 \sig(\mathcal{L})$.
Defining the harmonic oscillator annihilation operators $a_i = (r_i/l + l \partial_i)/\sqrt{2}$ (where $l=\sqrt{v/\kappa}$, $\omega = \sqrt{2v\kappa}$) we can write
\begin{equation}
    \mathcal{L} 
    =
    \begin{pmatrix}
        m\tau_0 & D \\
        D^\dag & -m\tau_0
    \end{pmatrix}
    \quad \text{where} \quad 
    D = \omega
    \begin{pmatrix}
        a_x & -a_y^\dag \\
        a_y & a_x^\dag
    \end{pmatrix}.
\end{equation}
The kernel of $D$ is spanned the two-dimensional harmonic oscillator ground state,
\begin{equation}
    \begin{pmatrix}
    e^{-(x^2+y^2)/(2l^2)}\\
    0
    \end{pmatrix}
    \in \ker D,
\end{equation}
while $\ker D^\dag=0$. This gives $\textrm{ind}(D)=+1$. Changing from positive to negative $m$ thus results in
\begin{align}
    \Delta \nu
    &= \frac12 \Delta \sig(\mathcal{L}) = \textrm{ind}(D) = +1. \nonumber
\end{align}
We again find that the localizer invariant follows that of the Hamiltonian.


\section{Spatial symmetries}\label{sec:crystalline_phases}

The previous section's discussion pertained to systems with \textit{on-site} symmetries, that is, symmetry that commute with the position operators $R_j$. 
We now turn to the case of phases protected by crystalline symmetries. 
The purpose of this section is to show how a generic crystalline symmetry of the Hamiltonian is inherited by the localizer. 
Once we know how the localizer inherits the crystalline symmetry it can be classified using the methodology of~\cref{sec:zero_dimensions}, and we complete the derivation of a classification scheme for systems with any real-space or on-site symmetry.

We consider a $d$-dimensional system with a crystalline symmetry $\mathcal{S}$, which may be any of the four types of `symmetry' discussed in \cref{sec:zero_dimensions}: unitary or anti-unitary, symmetry or anti-symmetry. 
We label the symmetry type using two quantities, $\alpha, \beta = \pm 1$. 
The first, $\alpha$, denotes whether it is a symmetry or anti-symmetry,
\begin{align}
    \mathcal S H\mathcal S^{-1} = \alpha H.
\end{align}
The second, $\beta$, distinguishes unitary and anti-unitary operators,
\begin{align}
    \mathcal S = \begin{cases}
        U   &\textup{for } \beta =+1, \\
        U \mathcal K &\textup{for } \beta =-1,
    \end{cases}
\end{align}
where $\mathcal K$ is complex conjugation.
The action of $\mathcal{S}$ on the position operators $R_j$ is always to enact a combination of rotation and reflection to the coordinate system,
\begin{align}
    \mathcal{S} R_j \mathcal{S}^{-1} = Q_{jk} R_k,
\end{align}
where $Q \in O(d)$ is the real-space representation of the orthogonal component of the symmetry on the position operators. Note that we continue to use implicit summation notation. 


We shall see how a symmetry of the Hamiltonian translates to a symmetry of the localizer. However, as in~\cref{sec:dim_recg_and_cliff}, the nature of the symmetry ($\alpha,\beta$) will change. 
The localizer of our $d$-dimensional system takes the form\footnote{If $H$ has chiral symmetry it can be written as $H = H_+\otimes\sigma_1+H_-\otimes\sigma_2$, and the localizer instead takes the form $\mathcal{L} = H_+ \otimes \Gamma_{D+1} + H_- \otimes \Gamma_{D+2} + \kappa R_j\otimes\Gamma_j$. The analysis below can then be carried over unchanged.}
\begin{align}
    \mathcal{L} = 
    H \otimes \Gamma_{D+1}
    +
    \kappa R_j \otimes \Gamma_j.
\end{align}
We wish to construct a symmetry $\widetilde{\mathcal{S}}$ of the localizer of the form
\begin{align}
    \widetilde{\mathcal{S}} = \mathcal{S} \otimes \Omega,
\end{align}
where $\Omega$ is an arbitrary operation acting on the localizer's auxiliary Clifford degrees of freedom. 
Our objective is to find an explicit form of $\Omega$ which ensures an $\widetilde{\mathcal{S}}$ that is a symmetry of~$\mathcal{L}$. 

Let us consider the action of $\widetilde{\mathcal{S}}$ on $\mathcal{L}$,
\begin{align}\label{eqn:l_rotated}
\begin{aligned}
    \widetilde{\mathcal{S}} \mathcal{L} \widetilde{\mathcal{S}}^{-1} 
    = 
    & (\alpha  H) \otimes \beta^{D+2} \big(\Omega\ \Gamma_{D+1} \Omega^{-1} \big) \\
    & + \kappa (Q_{jk} R_k) \otimes \beta^{j+1}\big(\Omega\ \Gamma_j\ \Omega^{-1} \big),
\end{aligned}
\end{align}
where $Q_{jk}$ emerges from the rotation of the position operators, $\alpha$ comes from the (anti)commutation of $\mathcal S$ with $H$. 
The factor $(\beta)^{j+1}=\pm1$ arises due to complex conjugation when $\mathcal{S}$ is anti-unitary, where we've chosen (without loss of generality) a representation of the Clifford algebra $\Gamma_j$ where 
\begin{align}
    \Gamma_j \in \begin{cases}
        \Re \textup{for $j$ odd},\\
        \Im \textup{for $j$ even}.
    \end{cases}
\end{align}
Though \cref{eqn:l_rotated} is reasonably involved, we can simplify it by defining a new orthogonal matrix $\widetilde{Q} \in O(d+1)$,
\begin{align}
    \widetilde{Q} =
    \begin{pmatrix}
    B Q & 0 \\
    0 & \alpha
\end{pmatrix}
\end{align}
where $B$ is a diagonal matrix of $\pm1$: $B_{jj} = \beta^{j+1}$.
\begin{align}\label{eqn:q_final}
\begin{aligned}
    \widetilde{\mathcal{S}} \mathcal{L} \widetilde{\mathcal{S}}^{-1} 
    = 
    & H \otimes  \big(\Omega \Gamma_{D+1} \Omega^{-1}\big) \widetilde{Q}_{D+1,D+1} \\
    & + \kappa R_k \otimes \big( \Omega \Gamma_j \Omega^{-1} \big) \widetilde{Q}_{jk},
\end{aligned}
\end{align}
where the implied sum in the second line only runs up to the dimensionality of the system $d$. 

In order for $\widetilde{\mathcal{S}}$ to be a symmetry of $\mathcal{L}$, \emph{i.e.} $\widetilde{\mathcal{S}} \mathcal{L} \widetilde{\mathcal{S}}^{-1} = \pm\mathcal{L}$, we must find a choice of $\Omega$ that removes the orthogonal rotation~$\widetilde Q$,
\begin{align}
    \Omega \Gamma_j \Omega^{-1} = \pm \Gamma_l \widetilde{Q}_{lj}^{-1}.
\end{align}
The two options $\pm 1$ on the RHS correspond to $\widetilde{\mathcal{S}}$ being either a symmetry or anti-symmetry of $\mathcal{L}$. 
Next, we explicitly construct $\Omega$ and show how whether $\widetilde{\mathcal{S}}$ is a symmetry or anti-symmetry depends on $\det \widetilde{Q} = \pm1$, \emph{i.e.} if $\widetilde{Q}\in SO(d+1)$ we will end up with a commuting symmetry, whereas if $\widetilde{Q} \notin SO(d+1)$ then we end up with an anticommuting symmetry. 

\subsection{Inherited symmetries}

We first construct $\Omega$ for the case of $\det \widetilde{Q} = +1$. 
Our aim is thus to lift the action \(\widetilde{Q}\in SO(d+1)\) on the Clifford generators to its corresponding spin representation \(\Omega\in \mathrm{Spin}(d+1)\)~\cite{Lawson_spin_1990, doran_geometric_2003}.
%
%
%
This can be achieved by first rewriting $\widetilde{Q}$ as a matrix exponent
\begin{equation}
    \widetilde{Q} = e^{\omega} \quad \textrm{where} \quad \omega = - \omega^T,
\end{equation}
which can then used to construct 
\begin{align}\label{eqn:clifford_rotor}
    \Omega = \exp \left (-\frac 14 \omega_{jk} \Gamma_j \Gamma_k\right ).
\end{align} 
In \cref{apx:n_dim_rotations} we show explicitly that this operator enacts the desired
\begin{align}
    \Omega \Gamma_j \Omega^{-1} = \Gamma_k \widetilde{Q}_{kj}^{-1},
\end{align}
for any $\widetilde{Q} \in SO(D+1)$. 

Thus, combining this with \cref{eqn:q_final}, we find that 
\begin{align}
    \widetilde{\mathcal{S}}\mathcal{L} {\widetilde{\mathcal{S}}}^{-1} = \mathcal{L}.
\end{align}

\subsection{Inherited anti-symmetries}



We next construct $\Omega$ for $\det \widetilde{Q} = -1$. 
As $\widetilde{Q}$ is no longer in $SO(D+1)$ we cannot directly reuse the above procedure to construct $\Omega$. 
We must instead first decompose
\begin{equation}
    \widetilde{Q} = A R    
\end{equation}
into the product of a proper rotation $\det R = +1$ and a reflection $\det A=-1$ through the hyperplane normal to $\hat{n}$. 
We similarly separate $\Omega$ into two components 
\begin{equation}
    \Omega = \Omega_R \Omega_A,     
\end{equation}
each enacting operations associated with $R$ and $A$ on the Clifford basis. 
The construction of $\Omega_R$ follows the same procedure as \cref{eqn:clifford_rotor}. 
Writing $R = e^\omega$, 
\begin{align}
    \Omega_R = \exp \left (-\frac 14 \omega_{jk} \Gamma_j \Gamma_k\right ).
\end{align} 
This enacts 
\begin{equation}
    \Omega_R \vec{\Gamma} \Omega_R^{-1} = \vec{\Gamma} R^{-1}.
\end{equation}

Choosing the reflection component, $\Omega_A$, as
\begin{align}
    \Omega_A = \hat{n} \cdot \vec{\Gamma},
\end{align}
we show, in \cref{apx:n_dim_reflections}, that conjugation with $\Omega_A$,
\begin{align}\label{eqn:clifford_reflector}
    \Omega_A \vec{\Gamma} \Omega_A^{-1} = - \vec{\Gamma} A,
\end{align}
leaves the component normal to the hyperplane unchanged and reverses all components within the hyperplane. 
We thus find that, unlike the case of proper rotations, the Clifford operation~\cref{eqn:clifford_reflector} also includes an overall factor of $-1$, which is why $\det \widetilde{Q} = -1$ gives rise to anti-symmetries of $\mathcal{L}$.
Together, the combined actions of $\Omega_R$ and $\Omega_A$ enact
\begin{align}
    \begin{aligned}
        \Omega \vec{\Gamma} \Omega^{-1} &= \Omega_R \Omega_A \vec{\Gamma} \Omega^{-1}_A \Omega^{-1}_R, \\ 
        & = - \Omega_R \vec{\Gamma} \Omega^{-1}_R A, \\
        & = - \vec{\Gamma} R^{-1} A \\
        & = - \vec{\Gamma} \widetilde{Q}^{-1},
    \end{aligned}
\end{align}
where we have used $A^{-1} = A$. Combining this with \cref{eqn:q_final}, we find that 
\begin{align}
    \widetilde{\mathcal{S}} \mathcal{L}\widetilde{\mathcal{S}}^{-1} &= -\mathcal{L}.
\end{align}

\section{Examples} \label{sec:examples}
Now that the method has been established, we construct a number of examples that illustrate the completeness and versatility of this procedure. Each time, the calculation is the same, we start with a chosen symmetry class and dimension and construct the hierarchy of localizers found by including different combinations of position operators. If the phase has obstructed atomic limits---which depend on translational symmetry in at least one dimension---we must also construct localizers using a partly Fourier-transformed Hamiltonian. For each localizer, we compute the symmetry class inherited from the symmetry of $H$. The exhaustive procedure for computing this is detailed in \cref{sec:crystalline_phases}, however, as we shall see, in most cases finding the symmetry of the localizer is simple enough to simply be guessed on inspection. Once the symmetry is understood, we break the localizer up into blocks and classify the 0D topological invariant for each block. 

An example is studied in each case, showing that the localizer reproduces, and sometimes extends the established classifications found in the literature. Remarkably, despite many of the momentum space invariants being highly involved, each time the localizer procedure is effectively identical, and generally rather simple. 

The structure of the examples is as follows. In \cref{sec:weak_and_weyl} we provide a simple example of a model hosting weak invariants and a Weyl semimetal phase. In \cref{sec:hoti_atomic_limit} we consider an inversion symmetric material that hosts higher-order topological phases and obstructed atomic limits, finding that the localizer can reproduce the wide variety of invariants---winding numbers and symmetry eigenvalues---that fully classify this system. This benchmarks our methodology against both previous localizer studies which use a slightly different construction \cite{cerjan_crystal_2024}, as well as ensuring it agrees with momentum-space invariants for crystalline phases.

\subsection{3d class A---Weak TIs and Weyl semimetals}
\label{sec:weak_and_weyl}

We consider a simple first example: a three-dimensional system in class A, with no imposed symmetries. This class hosts no strong phases in three dimensions. However, provided one assumes translational symmetry, the system can host both weak topological and Weyl semimetal phases. The conventional route to constructing a weak invariant starts by expressing the Hamiltonian in a partly Fourier-transformed basis, where  (without loss of generality) we apply the transformation in the $z$-direction, 
\begin{align}
\label{eq:FTr}
    H = \sum_{k_z} \ketbra{k_z}{k_z} \otimes H(k_z).
\end{align} 
If the Hamiltonian is gapped, we have the guarantee that all $H(k_z)$ are gapped, which ensures that they must all reside in the same topological phase. Thus, the weak invariant is found by choosing an arbitrary representative value of $k_z$, and computing the Chern number, effectively measuring whether the system is smoothly deformable to a stack of uncoupled Chern insulators. 

Following a similar procedure, we may diagnose a Weyl semimetal by computing the Chern number of $H(k_z)$ as a function of $k_z$. If there exist values of $k_z$ where the Chern number differs, then a Weyl node must exist between them where the spectrum of $H(k_z)$ becomes gapless \cite{Armitage2018}.

Diagnosing both of these cases is straightforward using the intermediate weak localizer $\mathcal{L}_{XY}(k_z)$ introduced~\cref{sec:multiple_loc},
\begin{equation}
\label{eqn:A_weyl_loc}
    \mathcal{L}_{XY}(k_z) = H(k_z)\, \sigma_3 + \kappa \left( X \sigma_1 + Y \sigma_2 \right).
\end{equation}
Its half signature probes the Chern number at $k_z$,
\begin{equation}
    \frac{1}{2} \sig \mathcal{L}_{XY}(k_z) = C(k_z),
\end{equation}
which identifies a weak phase so long as $\mathcal{L}_{XY}(k_z)$ is gapped for all $k_z$, and identifies a Weyl semimetal if $\mathcal{L}_{XY}(k_z)$ becomes gapless at a value of $k_z$ where the signature changes. For completeness we recall that, without Fourier transforming using \eqref{eq:FTr}, the Localizer spectrum of a Weyl semimetal exhibits zero-modes, separated from a continuum of states by a gap, whose number equals to the number of Weyl cones~\cite{schulz2022invariants,schulz2023spectral,franca2024topological,Franca2024_PRM}.

\begin{figure}[t]
    \centering
    \includegraphics[width=1\columnwidth]{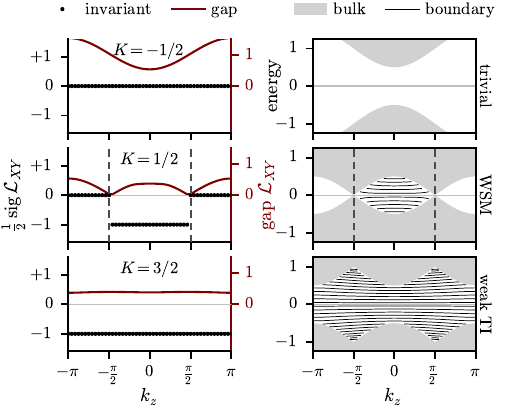}
    \caption{
    Localizer invariant and boundary spectrum for the three phases of Hamiltonian Eq.~\eqref{eqn:A_weyl_model}:
    $K=-1/2$ (trivial), $1/2$ (Weyl semimetal), $3/2$ (weak TI). 
    Left: half signature of $\mathcal{L}_{XY}=H\sigma_3+\kappa(X\sigma_1+Y\sigma_2)$ at $E=0$ (black) and its gap (maroon); the invariant probes the Chern number of the $k_z$ layer. 
    Right: projected bulk bands (grey) and the boundary states of the cross section open in $xy$ (black). 
    The system is $17\times17$ in the $xy-$plane with $k_z$ entering as a twisted boundary condition, and $\kappa=0.05$; dashed lines mark the Weyl nodes at $k_z=\pm\pi/2$.
    }
    \label{fig:A_3d_weyl}
\end{figure}

As an example, consider the following minimal two-band model \cite{chang_chiral_2015}, $H(k_z)$ given by,
\begin{align}
\label{eqn:A_weyl_model}
\begin{aligned}
    H(k_z)
    = 
    & \sum_{j = 1, 2}\, \sum_{\bm{r}} 
    \ketbra{\bm{r} + \hat{\bm e}_j}{\bm{r}}  
    \frac{\tau_3 + i \tau_j}{2} + \mathrm{h.c.}
    \\
    &+ m(k_z)  \sum_{\vec{r}} \ketbra{\bm{r}}{\bm{r}}\, \tau_3.
\end{aligned}
\end{align}
where $\bm{r}=(x,y)$, and $m(k_z) = K+ \tfrac12 \left(\cos k_z - 5 \right)$. 
The model is a trivial insulator for $K<0$, a Weyl-semimetal for $0<K<2$  and a weak TI for a weak-TI for $2<K<4$. These phases are easily distinguishable the localizer invariant, shown in Fig.~\ref{fig:A_3d_weyl}. 
The invariant is constant at $0$ ($-1$) for $K = -1/2$ ($K = 3/2$), with the localizer gap open for all $k_z$. 
At $K = 1/2$ the localizer invariant changes between $0$ and $-1$ at the Weyl-nodes, where the localizer gap closes. Note that this methodology, unlike the conventional route to an invariant that uses the Chern number, only depends on translational symmetry in the $z$-direction, and is still valid when the system has disorder in the $x, y$-directions.

\subsection{2d class AIII with \texorpdfstring{$\mathcal{I}$}{I}---Higher-order TIs and atomic limits}
\label{sec:hoti_atomic_limit}

Next, we consider an example that illustrates how to incorporate crystalline symmetries. While the localizer method has been shown to capture crystalline phases protected by an involutory phase (i.e.~the symmetry squares to the identity)~\cite{cerjan_crystal_2024}, the exact relation between localizer invariants and weak invariants has not been established. The following examples showcase how our method incorporates these cases, where the non-involutory case (i.e.~rotation symmetries) will be addressed in the next section (\cref{sec:ex_AII_C4}).

Let us consider a two-dimensional system in class AIII with inversion symmetry, where chiral and inversion anti-commute with one another $\{\mathcal C, \mathcal I \} = 0$. This symmetry class is sometimes denoted class $\textrm{AIII}^{\mathcal{I}_-}$ where the subscript in ${\mathcal{I}_-}$ labels this anti-commutation \cite{trifunovic2019higher,geier2018second}. Class $\textrm{AIII}^{\mathcal{I}_-}$ hosts a panoply of strong, weak, higher-order topological insulator (HOTI), and obstructed atomic limit (OAL) phases, and the conventional approach uses a combination of four symmetry indicator and two winding number invariants to fully classify them, leading to a $\mathbb{Z}^6$ classification. As we shall see, the localizer can capture them all.  

We review the conventional approach to classification. The symmetry indicators are computed by considering only the four points in the Brillouin zone preserved by inversion, these high-symmetry points are given by $\bm k^*$ at $\Gamma=(0,0), X=(\pi,0), Y=(0,\pi), M=(\pi,\pi)$. Writing the Bloch Hamiltonian in a basis where $\mathcal I = \tau_1$ and $\mathcal C = \tau_3$, the Hamiltonian at a high symmetry point takes the form
\begin{equation}
    H_{\bm k^*} 
    = 
    \begin{pmatrix}
        0 & A_{\bm k^*}  \\
        A_{\bm k^*} & 0    
    \end{pmatrix}, 
    \quad \textrm{where} \quad
    A_{\bm k^*} = A_{\bm k^*}^\dag.
\end{equation}
While $H_{\bm k^*}$ is constrained by chiral symmetry to have a symmetric spectrum, forcing $\sig H_{\bm k^*}=0$, $A_{\bm k^*}$ is not constrained, so we may construct an integer invariant at each high symmetry point by taking its signature,
\begin{equation}
    \nu(\bm k^*) = \frac12 \sig A_{\bm k^*} = \frac14 \sig \big( \mathcal{I} H_{\bm k^*} \big) \in \mathbb{Z}.
\end{equation}
This gives us four invariants, which we can rewrite by taking the following linear combinations (which will ensure they match exactly with the localizer based invariants computed later),
\begin{align}
    \nu_0 &= \nu(\Gamma), \\
    \nu_x &= \nu(\Gamma) + \nu(X), \\
    \nu_y &= \nu(\Gamma) + \nu(Y), \\
    \nu_{xy} &= \nu(\Gamma) + \nu(X) + \nu(Y) + \nu(M).
\end{align}
In addition to these symmetry indicators, two more invariants can be computed, found by taking a winding number that wraps the Brillouin zone in either the $k_x$ or $k_y$ direction. These are computed by setting an arbitrary value of $k_y$ ($k_x$), and computing the winding number in $k_x$ ($k_y$), 
\begin{align}
     W_{x}(k_y) &= \frac{1}{2\pi}\int d k_x \partial_{k_x} \log \det A_{\bm k}\in \mathbb Z, \label{eqn:wx}\\
     W_{y}(k_x) &= \frac{1}{2\pi}\int d k_y \partial_{k_y} \log \det A_{\bm k}\in \mathbb Z, \label{eqn:wy}
\end{align}
where the off diagonal block, $A_{\bm k}$, is not constrained to be Hermitian at generic $\bm k$. As we are considering gapped Hamiltonians only, any choice of the fixed momentum coordinate, i.e.~$k_y$ in \cref{eqn:wx} and $k_x$ in \cref{eqn:wy}, gives the same invariant. 

Thus, we have obtained the full $\mathbb Z^6$ classification. Note that of these invariants, $\nu_{xy}$ is the only strong (i.e.~manifestly 2D) invariant. When $\nu_{xy}$ is odd, the system forms a higher-order topological insulator with protected zero-energy corner states, whereas when it is odd the phase is an obstructed atomic limit \cite{trifunovic2019higher,shiozaki2014topology,geier2018second}. On the other hand, $\nu_x$, $\nu_y$, $W_x$ and $W_y$ are 1D weak invariants, classifying phases that can be realised with a stack of 1D systems, and $\nu_0$ classifies phases that can be realised with an array of 0D systems.

We now turn to how the localizer framework reproduces this classification and all associated invariants. We shall compute four localizers and derive all associated invariants: $H$ itself acts as the `trivial localizer' and can be used to measure $\nu_0$; $\mathcal L_X$ includes the $X$ position operator and will compute $\nu_x$ and $W_x$; ; $\mathcal L_Y$ includes the $Y$ position operator and will compute $\nu_y$ and $W_y$; finally, $\mathcal L_{XY}$ includes both position operator and will compute $\nu_xy$ invariant. Let us consider each in turn.  

To compute $\nu_0$ we take advantage of the unitary inversion symmetry to Block-diagonalise the Hamiltonian into two blocks, $H_+$ and $H_-$. Chiral symmetry anti-commutes $\mathcal I$ so it exchanges these blocks. It also anti-commutes with $H$, so the blocks must differ by a sign. Thus, we can write the Hamiltonian in the form,  
\begin{align} \label{eqn:0DclassAIII}
    H = \begin{pmatrix}
        H' & \\ & -H'
    \end{pmatrix},\textup{ with }\mathcal I = \tau_3\textup{ and }\mathcal C = \tau_1.
\end{align}
Here, the submatrix $H'$ is in class A, with neither chiral nor inversion symmetry. In order to access the 0D weak invariant---which depends on translational symmetry---we Fourier transform $H'$ in both $x$ and $y$, choose a representative momentum point, e.g.~$\Gamma$---the choice does not matter since the phase is gapped everywhere, and take the signature, 
\begin{equation}
    \nu_0 = \frac12 \sig (H'_\Gamma) = \frac14 \sig (\mathcal{I}H_\Gamma).
\end{equation}

Next, we consider the 1D invariants, $\nu_x$, $W_x$ and $\nu_y$, $W_y$. We will detail the procedure for recovering the two $x$-invariants, where the other two are found via an identical computation with $x\leftrightarrow y$. First we take a Fourier transform of the Hamiltonian in the $k_y$ direction, leaving $x$ in real space and obtaining $H(k_y)$. As before, the phase is gapped so we may choose a representative $k_y$ value. The $\mathcal L_X$ localizer is then calculated by following the procedure in \cref{sec:one_step} for a chiral $H$,  
\begin{align} \label{eqn:lx_ky}
    \mathcal L_{X,k_y} = H_{k_y} + \kappa \mathcal C X. 
\end{align}
This localizer breaks the original $\mathcal C$ symmetry, however if $k_y$ is at a high-symmetry point $k_{y}^*$ (i.e.~0 or $\pi$) it recovers $\mathcal I$, since $\{\mathcal I, X\} = 0 \implies [\mathcal I, \mathcal C X] = 0$. Here, we can split the localizer into two blocks, indexed by the eigenvalue of $\mathcal I$, 
\begin{align}
    \mathcal L_{X,k^*_{y}} = \begin{pmatrix}
       \mathcal L_{X,k^*_{y}}^+ & \\ & \mathcal L_{X,k^*_{y}}^-
    \end{pmatrix}.
\end{align}
Each block has no remaining symmetry, so we may classify them by their signature, arriving at two $\mathbb Z$ invariants, which we can now connect directly to $\nu_x$ and $W_x$ in the following combination,
\begin{align}
    W_x = \frac 12 \sig \mathcal L_{X,k_y}
\end{align}
which can be computed at any value of $k_y$, and 
\begin{align}\begin{aligned}
    \nu_x &= \frac 12 \sig  (\mathcal I \mathcal L_{X,k^*_{y}} )\\
    &= \frac 12 \left (\sig \mathcal L_{X,k^*_{y}}^+ -\sig \mathcal L_{X,k^*_{y}}^-\right ),
\end{aligned}
\end{align}
which must be computed at a high-symmetry $k_y$.

Thus, including the two invariants found from considering $\mathcal L_Y$, we recover the four 1D symmetry indicator and winding number invariants. 

Finally, we consider the $\mathcal L_{XY}$ localizer, which uses the full $H$ in real space. Following the procedure in \cref{sec:one_step} twice, we include the $X$ operator using the chiral symmetry, analogous to \cref{eqn:lx_ky} and then expand the Hilbert space to include the $Y$ operator,
\begin{align}
    \mathcal L_{XY} = (H + \kappa \mathcal C X) \sigma_1 + \kappa Y \sigma_2.
\end{align}
This has a chiral symmetry, given by the absent Clifford matrix, $\widetilde {\mathcal C} = \sigma_3$. In addition, we can write down an expanded inversion operator, following \cref{sec:crystalline_phases}, which takes the form
\begin{align}
    \widetilde{\mathcal I} = \mathcal I \sigma_1,
\end{align}
which also anti-commutes with $\widetilde{\mathcal C}$, such that this localizer is also in class $\textrm{AIII}^{\mathcal{I}_-}$. This means that we can follow a completely identical procedure to the discussion surrounding \cref{eqn:0DclassAIII}, since the class of $H$ and $L_{XY}$ is the same. We use the inversion and chiral symmetry to Block diagonalise the localizer,
\begin{align}
    \mathcal L_{XY} = \begin{pmatrix}
        \mathcal L_{XY}' & \\ & -\mathcal L_{XY}'
    \end{pmatrix}
\end{align}
and take the signature
\begin{align}
    \nu_{xy} = \frac12 \sig (\mathcal L_{XY}') = \frac14 \sig \left (\widetilde{\mathcal I}\mathcal L_{XY}\right ).
\end{align}
Thus, we have recovered all six momentum-space invariants using the localizer procedure.

\begin{figure}[t]
    \centering
    \includegraphics[width=1\columnwidth]{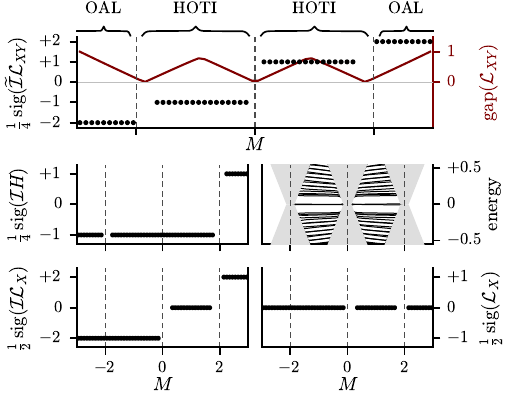}
    \caption{
    Localizer invariants of the Hamiltonian in Eq.~\cref{eqn:AIII_inversion_model} across its four gapped phases (braces).
    Top: Strong invariant $\tfrac14\sig(\widetilde{\mathcal{I}}\mathcal{L}_{XY})$ (black) and the localizer gap (maroon).
    Below: the remaining weak invariants, and the spectrum --- the states in open boundary conditions (black) overlayed with the bulk spectrum (grey).
    The system is $23\times23$ with periodic boundary conditions (wires of $23$ sites for $\mathcal{L}_{X}$), $\kappa = 0.27$, and $m = 0.1$; dashed lines mark the transitions of the bare model at $M = -2, 0, 2$.
    Localizer invariants are masked where the corresponding localizer's gap is below $0.1$. 
    We did not plot the $\mathbb{Z}^2$ invariants from $\mathcal{L}_Y$ as they are, for this particular model, the same as $\mathcal{L}_X$ (the model has an accidental $C_4$).
    }
    \label{fig:AIII_inv_2d}
\end{figure}

Next we consider an example to demonstrate how the Localizer invariants correspond to the BZ invariants we laid out above. 
We consider the following toy model given by the Hamiltonian,
\begin{align}\label{eqn:AIII_inversion_model}
    \begin{aligned}
    H  = &\sum_{j = 1,2}\sum_{\bm r} \ketbra{\bm{r}+ \hat {\bm e}_j}{\bm{r}} 
    \frac{i \tau_j \sigma_2-\tau_0 \sigma_1}{2} + h.c.
    \\ 
    & + \sum_{\bm r} \ketbra{\bm{r}}{\bm{r}}\left [ M \tau_0\sigma_1 
    + m (\tau_1 + \tau_2)\sigma_1\right ]
\end{aligned}
\end{align}
with chiral symmetry $C = \tau_0 \sigma_3$ and inversion $\mathcal{I} = \tau_0 \sigma_1$.
Here $m$ is a mass term: $\tau_2\sigma_1$ gaps the $x$-normal edges and $\tau_1\sigma_1$ the $y$-normal ones (in the HOTI phase). 
$M$ is a bulk mass term that drives the model between four different bulk phases, with transitions at $M = -2$, $0$, and $2$ (when $m=0$).
The four gapped phases carry $(\nu_0, \nu_x, \nu_{xy})$ of
\begin{align*}
 (-1, -2, -2) \quad &\textrm{for}\quad M < -2, \\
 (-1, -2, -1) \quad &\textrm{for}\quad -2 < M < 0, \\
 (-1,  0, +1) \quad &\textrm{for}\quad 0 < M < 2, \\
 (+1, +2, +2) \quad &\textrm{for}\quad M > 2,
\end{align*}
where we have omitted $\nu_y$ as $\nu_y=\nu_x$ for this model, and the windings are $W_x=W_y=0$ trivial in all phases. 
Only $\nu_{xy}$ distinguishes all four phases. 
These values are all in agreement with the localizer results of Fig.~\ref{fig:AIII_inv_2d}. 

\subsection{2D class AII with \texorpdfstring{$C_4$}{C4}---Beyond symmetry indicators}\label{sec:ex_AII_C4}

\begin{figure}[t]
    \centering
    \includegraphics[width=1\columnwidth]{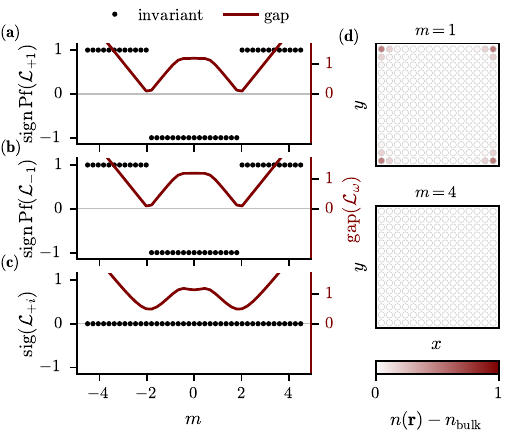}
    \caption{
    (a)-(c) Localizer invariants per $\tilde{C}_4$ eigenspace as a function of $m$: the Pfaffian signs $\mathrm{sign,Pf},\mathcal{L}_{\pm 1}$ of the $\omega = \pm 1$ (class D) blocks and the signature $\mathrm{sig},\mathcal{L}_{+i}$ of the $\omega = +i$ (class A) block (dots), together with each block's spectral gap (solid lines), for an open $11\times11$ lattice with $\kappa = 0.14$ and $\Delta = 0.6$. (d) Deviation of the site-resolved density of the occupied states from its bulk value for $m = 1$, where charge accumulates at the corners, and $m = 4$, where it is uniform.
    }
    \label{fig:AII_C4_2d}
\end{figure}

Next, let us consider a two dimensional system which has both $\mathcal T^2 = -1$ and a spinful $C_4$ rotation symmetry, with $(C_4 \mathcal T)^4 = -1$. This symmetry class was shown under an extensive $K$-theoretic study to host a $\mathbb{Z}_2^2 \times \mathbb{Z}$ classification \cite{cornfeld2019}, however the study did not provide invariants. Later, an extensive approach using symmetry indicators was able to construct an invariant that distinguished the $\mathbb Z$ component but the $\mathbb Z_2^2$ remained elusive \cite{schindler2019}. Since then, a highly involved symmetry indicator invariant was proposed for the remaining component \cite{kooi2021bulk}, which we do not reproduce here due to space constraints. Finally, within the spectral localizer literature, no invariant has been constructed for this phase, falling outside the symmetry classes considered in Ref.~\cite{cerjan_crystal_2024}. Here we show that writing this invariant is not only possible with our formalism, but also remarkably straightforward to compute.

We start by classifying the symmetries of the system. Time-reversal takes the usual spinful form, $\mathcal T = is_y\mathcal K$, and the $C_4$ rotation operator includes a spin-rotation component $e^{-i \pi s_z/4}$ which ensures that $C_4^4 = -1$. The localizer is straightforwardly computed using the procedure in \cref{sec:dim_recg_and_cliff}, given by
\begin{align}
    \mathcal L_{XY} = H \sigma_3 + \kappa X  \sigma_1 + \kappa Y \sigma_2. 
\end{align}
Each symmetry of the Hamiltonian leads to a symmetry of the localizer. Following the Bott clock procedure shown in \cref{sec:loc_tenfold}, we see that the $\mathcal T$ symmetry of the Hamiltonian becomes a particle hole symmetry of the localizer, given by
\begin{align}
    \widetilde {\mathcal P} = \mathcal T \sigma_2,\textup{ with } \widetilde {\mathcal P}^2 = +1,
\end{align}
which anti-commutes with the localizer. The $C_4$ symmetry is less straightforward to compute, since it involves both a real space and spin component. Na\"ively applying this symmetry to the localizer yields an operator with $X$ and $Y$ swapped,
\begin{align}
    C_4 \mathcal L_{XY} C_4^{-1} = H \sigma_3 + \kappa Y  \sigma_1  -\kappa X \sigma_2. 
\end{align}
This, however, can be fixed by including a $\sigma$-space component that reverses the rotation of the position coordinates by swapping our Pauli matrices, following exactly the procedure outlined in \cref{sec:crystalline_phases}, 
\begin{align}
    \widetilde C_4 = C_4 e^{-i\frac \pi 4 \sigma_3}, 
\end{align}
which commutes with $\mathcal L_{XY}$ and satisfies $\widetilde {C}_4^4 = +1$. The eigenvalues of $\widetilde {C}_4$ are $\omega \in \{1,-1,i,-1\}$, and we may use this to block diagonalise the localizer, finding a basis in which 
\begin{equation}\label{eqn:AII_C4_and_t_loc}
    \mathcal L_{XY} =
    \begin{pmatrix}
        \mathcal{L}_{XY}^{+1} &&&\\    
        & \mathcal{L}_{XY}^{-1} &&\\    
        && \mathcal{L}_{XY}^{+i} &\\    
        &&& \mathcal{L}_{XY}^{-i}
    \end{pmatrix}.
\end{equation}
Since $\widetilde{\mathcal P}$ commutes with $\widetilde C_4$, the effect of $\widetilde{\mathcal P}$ is to apply complex conjugation to the eigenvalues of $\widetilde C_4$. This means that $ \mathcal{L}_{XY}^{\pm1}$ are left intact under $\widetilde {\mathcal P}$---and so have PHS---whereas $ \mathcal{L}_{XY}^{\pm i}$ are exchanged under it and so individually have no symmetry, although they are forced to have opposite signatures. 
\begin{align*}
    \mathcal{L}_{\pm 1} \in \textrm{class D}, \\
    \mathcal{L}_{\pm i} \in \textrm{class A}.
\end{align*}
Thus, we may write the invariants by taking
\begin{align}
    \sgn \textrm{Pf}\ \mathcal{L}_{XY}^{+1} &\in \mathbb Z_2, \\ 
    \sgn \textrm{Pf}\ \mathcal{L}_{XY}^{-1} &\in \mathbb Z_2, \\ 
    \sig \mathcal{L}_{XY}^{+i} &\in \mathbb Z,
\end{align}
with $\sig \mathcal{L}_{XY}^{+i} = - \sig \mathcal{L}_{XY}^{+i}$. Thus, we recover the full $\mathbb{Z}^2_2 \times \mathbb Z$ classification.
The Pfaffian of $\mathcal{L}$ diagnoses the presence of a quantum spin-Hall phase (QSH), independent of the crystalline symmetry, while the Pfaffians of the blocks $\mathcal{L}_{\pm1}$ diagnose QSH phases that are distinct in the presence of $C_4$. Phases with Pfaffian $(1,1)$ are atomic limits with an anomalous\footnote{A corner-charge of $\pm1$ is anomalous because Kramer's degeneracy forces corner decorations to have even charge.} corner-charge of $\pm1$. The remaining $\mathbb{Z}$ distinguishes between other obstructed atomic limit phases.

As an example, consider the model from Ref.~\cite{song2017}, which we re-express in real space as
\begin{align}\begin{aligned}
    H &=  
    \sum_{\bm r} \ketbra{\bm r + \hat{\bm e}_x }{\bm r}
    \frac {i \tau_0 \gamma_1 s_1 - h_-s_0} 2 + h.c.\\
    &+ \sum_{\bm r} \ketbra{\bm r + \hat{\bm e}_y }{\bm r}
    \frac {i \tau_0 \gamma_1 s_2 - h_+s_0} 2 + h.c.\\
    &+ \sum_{\bm r} \ketbra{\bm r }{\bm r} m\, \tau_0 \gamma_3 s_0
\end{aligned}
\end{align}
with $h_\pm = \tau_0 \gamma_3 \pm \Delta \tau_2 \gamma_2$ and time-reversal and fourfold-rotation symmetries
\begin{equation}
    \mathcal{T} = -i s_2 K, \qquad C_4 = \tau_3 e^{-i\pi s_3/4}.
\end{equation}
This model supports two phases with $\Delta = 0.6$. 
When $|m|>2$ it's an atomic limit with two Kramer's pairs of Wannier states around Wyckoff position $1a = (0,0)$  with $C_4$ eigvals $\pm \pi/4$ and $\pm 3\pi/4$.
When $|m|<2$ the Kramers pairs sit at $1b = (1/2,1/2)$ and the phase host and anomalous corner charges of $Q=1\mod2$. 
These phases cannot be distinguished using symmetry indicators as their occupied bands have the same set of $C_4$ eigenvalues at high-symmetry momenta.

We can, however, distinguish these two phases with the localizer invariant. The $|m|>2$ and $|m|<2$ phases come with localizer invariants $(+,+,0)$ and $(-,-,0)$, respectively.
Results are shown in~\cref{fig:AII_C4_2d}.

\subsection{\texorpdfstring{$C_4\mathcal T$}{C4T} obstructed atomic limit}
\label{sec:ex_C4T}

Next, we consider a model similar to that in \cref{sec:ex_AII_C4}, however where we break both $C_4$ and $\mathcal T$, but keep their product $C_4\mathcal T$ with $(C_4\mathcal T)^4 = -1$
All the topological phases hosted by this symmetry class are atomic limits, where none of them have any anomalous boundary states, though they may host anomalous boundary charges. This means that these phases cannot be distinguished using scattering invariants \cite{fulga2012scattering}. From $K$-theory studies, this symmetry class was shown to host a strong $\mathbb{Z}_2$ classification~\cite{cornfeld2019,shiozaki2023generalized}. 
This symmetry class has a stable $\mathbb{Z}_2$ classification~\cite{cornfeld2019,shiozaki2023generalized}. If we further impose translation symmetries, we instead get a $\mathbb{Z}_2\times\mathbb{Z}$ classification \cite{serrano2025magnetic}.\footnote{Note that this classifying group is sometimes given with an extra $\mathbb{Z}$~\cite{serrano2025magnetic}. This is simply the number of occupied states of $H$.} However, to our knowledge, no explicit invariant has been constructed in either real or momentum space that can capture the full $\mathbb{Z}_2\times\mathbb{Z}$ classification. 

A $\nu_4\in \mathbb Z_4$ invariant was constructed that differentiates atomic limits by their four different possible anomalous corner charges \cite{araya-day_pfaffian_2023}, which we shall later show is a quotient group of the full $\mathbb Z_2\times \mathbb Z$. The $\mathbb Z_4$ invariant is itself reasonably involved, found by taking the Berry flux through a quarter of the Brillouin zone. Corrections must be included for the phase of the open Wilson line from $\Gamma$ to $M$, gauge fixed by the Pfaffians of the $C_4\mathcal T$ overlap matrix at its end points. The full expression obtained is 
\begin{equation} \label{eqn:araya_invariant}
    \nu_4 = \frac{1}{\pi}\left[\int_{\mathrm{IBZ}} \mathrm{tr}\,\mathcal{F}\, d^2k
    + 2\,\mathrm{Im}\log \widetilde{\det}\, W_{\Gamma\to M}\right] \bmod 4,
\end{equation}
where $\mathcal F$ is the non-Abelian Berry curvature, $\widetilde{\det} W_{\Gamma\to M}$ is a dressed Wilson line determinant. The details of the computation are found in Ref.~\cite{araya-day_pfaffian_2023}.

We now show how the localizer procedure may be used to compute this invariant. Two localizers will be constructed, a weak localizer $\mathcal L_X$ which gives the $\mathbb Z$ component, and an $\mathcal L_{XY}$ localizer that contributes the $\mathbb Z_2$ component. The system is reasonably similar to that considered in \cref{sec:ex_AII_C4}, and so we may reuse the same operators derived there. Let us start by writing down the strong $\mathbb Z_2$ invariant, which is computed from 
\begin{align} \label{eqn:A_C4T_loc}
    \mathcal{L} = H \sigma_3 + \kappa X \sigma_1 + \kappa Y\sigma_2,
\end{align}
where $\sigma_i$ are Pauli matrices. This Hamiltonian breaks both the $\mathcal T$ and $C_4$ operators given in the previous section, which means that this localizer does not respect the $\widetilde {\mathcal P}$ and $\widetilde C_4$. However, the Hamiltonian respects $C_4\mathcal T$, so the localizer respects the product of $\widetilde {\mathcal P}$ and $\widetilde C_4$,
\begin{align}
    \widetilde{C}_4 \widetilde {\mathcal P} =  C_4 \mathcal T e^{i\frac \pi 4 \sigma_3} \sigma_2, 
\end{align}
which satisfies $\{\widetilde{C}_4 \widetilde {\mathcal P}, \mathcal L_{XY}\} = 0$ and $(\widetilde{C}_4 \widetilde {\mathcal P})^4 = +1$. In addition, we may construct a unitary effective $\widetilde C_2$ operator given by,
\begin{align} \label{eqn:c2}
    \widetilde C_2 = (\widetilde{C}_4 \widetilde {\mathcal P})^2 = (C_4\mathcal T)^2 \sigma_3,
\end{align}
which has eigenvalues $\pm 1$. This is then used to block diagonalise the localizer, 
\begin{equation}\label{eqn:AII_C4_loc}
    \mathcal L_{XY} \equiv
    \begin{pmatrix}
        \mathcal L_{XY}^+ &\\    
        & \mathcal L_{XY}^-    
    \end{pmatrix}.
\end{equation}
In each sector, $\widetilde{C}_4 \widetilde{\mathcal{P}}$ now acts like a conventional AZ particle hole symmetry, such that we can write down the symmetry class in each sector,
    \begin{align*}
        \mathcal L_{XY}^+: (\widetilde{C}_4 \widetilde{\mathcal{P}})^2 = +1 \implies \textrm{ class D}, \\
        \mathcal L_{XY}^- : (\widetilde{C}_4 \widetilde{\mathcal{P}})^2 = -1 \implies \textrm{ class C}.
    \end{align*}
Since class C has no zero-dimensional invariant, whereas class D has a $\mathbb Z_2$ invariant, so we find an overall $\mathbb{Z}_2$ classification, fopund by computing only the topology of $\mathcal L_{XY}^+$ block, 
\begin{equation}\label{eqn:C4T_Z2}
    \nu_{XY} = \sgn \textrm{Pf}\ \mathcal{L}_{+} \in \mathbb{Z}_2.
\end{equation}

To obtain the remaining $\mathbb{Z}$ classification---computed analogously to the weak invariants in \cref{sec:weak_and_weyl} and depends on translational symmetry to label a OAL---we demand translational symmetry in the $y$-direction. This allows us to construct a partially Fourier-transformed Bloch Hamiltonian, $H_{k_y}$, which is then be used to write a localizer,
\begin{align}
    \mathcal L_{X,k_y} = H_{k_y}\sigma_3 + \kappa X\sigma_1. 
\end{align}
This has a chiral symmetry, $\widetilde{\mathcal C} = \sigma_2$. When $k_y$ sits at a high-symmetry point (i.e.~$k_y^*=$ 0 or $\pi$), the localizer also gains a unitary $\widetilde C_2$ symmetry, \cref{eqn:c2}, which anti-commutes with the chiral symmetry $\{ \widetilde{\mathcal C},\widetilde C_2\}=0$. This means we can block diagonalise $\mathcal L_{XY}$ into two components which are exchanged by chiral symmetry, and themselves have no symmetries, falling into class A,
\begin{align}
    \mathcal L_{X,k_y^*} = \begin{pmatrix}
    \mathcal L_{X,k_y^*}^+ & \\ & \mathcal L_{X,k_y^*}^-
    \end{pmatrix}. 
\end{align}
This localizer is the straightforwardly computed as, 
\begin{equation}\label{eqn:C4T_Z}
    \begin{aligned}
        \nu_X &= \frac12 \sig L_{X,k_y^*}^+ \in \mathbb{Z}, \\ 
        &= \frac14 \sig\left( \widetilde{C}_2 \mathcal{L}_X \right).
    \end{aligned}
\end{equation}

\begin{figure}[t]
    \centering
    \includegraphics[width=1\columnwidth]{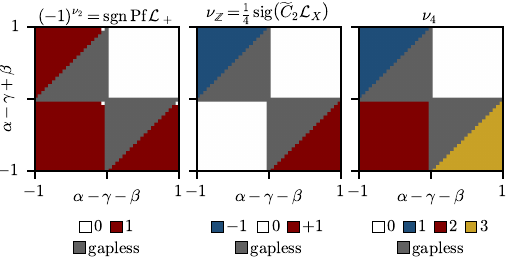}
    \caption{
        Localizer invariants of the $C_4\mathcal{T}$ model across its phase diagram, in the plane $(\alpha-\gamma-\beta,\ \alpha-\gamma+\beta)$ with $\gamma = 1-\alpha-|\beta|$.  
        The four corners of the phase diagram are the representative models of the four phases. 
        Left: the $\nu_2 \in \mathbb{Z}_2$ localizer invariant from an open $11\times 11$ sample with $\kappa = 0.2$. 
        Middle: the $\nu_\mathbb{Z}\in\mathbb{Z}$ localizer invariant from a $41$ site wire at $k_y = 0$ with $\kappa = 0.05$. 
        Right: the $\nu_4\in\mathbb{Z}_4$ Brillouin-zone invariant from Ref.~\cite{araya-day_pfaffian_2023} for comparison. 
        Grey marks gapless regions, where the bulk gap closes at Dirac points. 
    }    
    \label{fig:C4T}
\end{figure}

As an example, we consider the model used in~\cite{araya-day_pfaffian_2023}, demonstrating that the localizer is able to distinguish all four phases that were found using \cref{eqn:araya_invariant}.  The model consists of a single atom per square unit cell carrying a spin and $p_x, p_y$ orbital degree of freedom, with $\tau_i$ acting on the orbital and $s_i$ on the spin degree of freedom. 
It interpolates between three atomic limits,
\begin{equation}\label{eqn:C4T_model}
    H = \alpha H^{(0)} + \beta H^{(1)} + \gamma H^{(2)},
\end{equation}  
where $\gamma = 1 - \alpha - |\beta|$. The representative models being interpolated between are
\begin{align}
    H^{(0)} &= \sum_{\bm r} \ketbra{\bm r}{\bm r} \tau_3 s_3, \\
    H^{(1)} &= \sum_{\bm r} \ketbra{\bm r + \hat{\bm e}_x}{\bm r} T_{+}\frac{s_3 - is_1}2 +h.c.\nonumber \\
    &-\sum_{\bm r} \ketbra{\bm r + \hat{\bm e}_y}{\bm r} T_{-}\frac{s_3 - is_2}2 +h.c.\\ 
    H^{(2)} &= \sum_{\bm r} \ketbra{\bm r + \hat{\bm e}_x + \hat{\bm e}_y}{\bm r} T_{+}\frac{s_3 - is_1}2 +h.c.\nonumber\\
    &- \sum_{\bm r} \ketbra{\bm r - \hat{\bm e}_x + \hat{\bm e}_y}{\bm r} T_{+}\frac{s_3 - is_2}2 +h.c.
\end{align}
where $T_\pm = (\tau_0 \pm \tau_3) /2$. Each $H^{(\nu)}$ has two flat bands at $\pm 1$ with the occupied states sitting at different atomic limits,
\begin{itemize}[itemsep=2pt, topsep=5pt]
    \item at $1a = (0,0)$ for $\nu = 0$,
    \item at $1b = (1/2,1/2)$ for $\nu = 2$,
    \item at $2c = (1/2,0)$ and $(0,1/2)$ for $\nu = 1$ and $\nu = 3$,
\end{itemize}
We compute our localizer invariant $(\nu_{XY},\nu_X)$, for the model~\cref{eqn:C4T_model}. We contrast this with the $\nu_4$ from~\cite{araya-day_pfaffian_2023} in~\cref{fig:C4T}. The localizer invariant is able to distinguish between all four phases, with the $\mathbb{Z}_4$ emerging as a quotient of the full classifying group
\begin{equation}
    \nu_4 = 2\nu_{XY} + \nu_X\mod 4.
\end{equation}

\section{Conclusions}

We have presented a methodology for constructing simple, numerically efficient invariants for systems with an arbitrary set of unitary and antiunitary symmetries acting in real space. The procedure is based on classifying the spectral localizer operator as if it were a 0D Hamiltonian, constructed from a set of position operators and the Hamiltonian matrix in real-space. We show how the symmetry class of the spectral localizer descends from the symmetry class of the original Hamiltonian, which defines what invariant can be calculated. The 0D invariants, which are either signatures, or (the sign of) determinants and Pfaffians, are numerically efficient and mathematically simple topological invariants for all non-interacting classes.

We have given a set of examples for which a subset of existing methods to compute invariants fail to diagnose or for which a complete invariant did not exist. Specifically, we have constructed spectral localizer invariants for weak phases and phases protected by rotational symmetries, which was previously not captured by the localizer formalism. We have also discussed examples of atomic limits which were previously inaccessible with scattering or localizer invariants, providing a simpler, numerically efficient invariant for a phase with $C_4$ and $\mathcal{T}$ symmetries, simplifying previous results~\cite{kooi2021bulk}. 

We see no obstruction in applying this procedure to other scenarios where topological phases exist. An interesting direction is to include periodically driven (Floquet) systems, e.g. by extending previous spectral-localizer methods to include crystalline symmetries~\cite{Ghosh2024floquet}. 
A related appealing direction is to build on the known classifications of quadratic Lindbladians ~\cite{Lieu2019,Altland2020,Sa2022,Kawabata2022,Mao2023} to extend the spectral localizer construction to open quantum systems.
It would be fruitful to extend our discussion to non-symmorphic symmetries, as well as approximate and statistical symmetries~\cite{fulga2014statistical,chaou2025,chen2025intrinsic,zijderveld2026symmetric} through the symmetric approximant formalism discussed in Ref.~\cite{zijderveld2026symmetric}. Lastly, it would be desirable to extend our formalism to interacting symmetry-protected topological phases.

\textit{Acknowledgments---} We are grateful to Isidora Araya Day, Anton Akhmerov, Andres Perez Fadon, Frank Schindler, Ryan Barnett and Alex Cerjan for useful discussions.
All authors are supported by the European Research Council (ERC) Consolidator grant under grant agreement No. 101042707 (TOPOMORPH).

\bibliography{refs.bib}


\appendix
\crefalias{section}{appendix}
\crefalias{subsection}{appendix}

\section{Identities for the Clifford algebra and Pauli matrices}

A representation of the Clifford algebra is a set of matrices, $\Gamma_j$ which satisfy the following commutation relations,
\begin{align}
    \{
    \Gamma_{j},\Gamma_k
    \} = 2\delta_{jk}\1,
\end{align}
which implies the following commutation relations
\begin{align}
    [\Gamma_{j},\Gamma_k
    ] = 2(\Gamma_j\Gamma_k-\delta_{jk}).
\end{align}
The Pauli matrices,
\begin{align}
    \sigma_x = \begin{pmatrix}
        &1\\1&
    \end{pmatrix}, \,
    \sigma_y = \begin{pmatrix}
        &-i\\i&
    \end{pmatrix}, \,
    \sigma_z = \begin{pmatrix}
        1&\\&-1
    \end{pmatrix},
\end{align}
provide a 3-element representation of the Clifford algebra.

\subsection{New Clifford representations from old} \label{apx:pauli_building}
Suppose we have an $n$-element irrep of the Clifford algebra $\{\Gamma_j\}$, then there is a systematic way to construct an $n+2$ irrep of the Clifford algebra using the Pauli matrices, which we denote with $\Gamma_j'$ in the following way,
\begin{align}
    \Gamma_j' &= \sigma_x \otimes \Gamma_j\, \forall j \leq n, \\
    \Gamma_{n+1}' &= \sigma_y \otimes \1, \\ 
    \Gamma_{n+2}' &= \sigma_z \otimes \1.
\end{align}
Thus, we can start with the Pauli representation, and use this construction recursively to create an irrep with any (odd) number of elements. 

Any representation of the Clifford algebra built using this procedure has the property 
\begin{align}
    \Gamma_{n} \in \begin{cases}
 \Re & \textup{for } n \textup{ odd},
 \\ 
 \Im & \textup{for } n \textup{ even}.
\end{cases}
\end{align}

\subsection{Commutation identities}

Writing a commutator as $[ A, B]_-$ and an anticommutator as $[A, B]_+$, we can derive the following identity,
\begin{align}\begin{aligned}
    [A\otimes a, B\otimes b ]_{\pm} = \frac 12 & \left ( 
       [A,B]_+ \otimes [a,b]_\pm \right .\\
    & \left . + [A,B]_- \otimes [a,b]_\mp \right ).
\end{aligned}
\end{align}
Now we consider the case where $A,B$ are elements of the Clifford algebra, $\Gamma_j$ and $\Gamma_k$,
\begin{align}
    [\Gamma_j \otimes a, \Gamma_k\otimes b ]_{\pm} = 
    \begin{cases*}
    \1 \otimes [a,b]_{\pm}& for \;$j=k$ \\
    \Gamma_j \Gamma_k \otimes [a,b]_{\mp}& for \;$j\neq k$
    \end{cases*}.
\end{align}

\subsection{N-dimensional rotations} \label{apx:n_dim_rotations}

Let us consider a rotation matrix in $D$ dimensions, $R \in SO(D)$. Additionally, we consider a matrix that is defined by a vector $\bm v \in \mathbb R^D$ using a representation of the Clifford algebra $\Gamma_j$, as
\begin{align}
    \slashed { v} = v_j \Gamma_j.
\end{align}
Under the action of the rotation operator, the vector transforms as 
\begin{align}
    v'_j = R_{jk}v_k.
\end{align}
Our task is to find a matrix $\Omega$ which acts on the Clifford basis such that it recreates the action of the rotation $R$ on $\bm v$,
\begin{align}\begin{aligned}
    \Omega \slashed { v} \Omega^{-1} &= v_j\, \Omega \Gamma_j \Omega^{-1}, \\
     &= v_j\, R_{jk}^{-1} \Gamma_k ,\\
    &= \slashed { v}' .
\end{aligned}
\end{align}
For experts, this amounts to elucidating the connection between the groups $SO(n)$ and $\textup{Spin}(n)$ \cite{Lawson_spin_1990}.

In order to do this, first we must take the matrix logarithm of $R$ to obtain the antisymmetric matrix $\omega \in \mathfrak{so(D)}$,
\begin{align}\label{eqn:log_R}
    \omega = \log(R). 
\end{align}
Next, we construct the Clifford element,
\begin{align}
    B = \frac{\omega_{jk}}{4}\left [\Gamma_j, \Gamma_k\right ],
\end{align}
where the operator $\Omega$ is given by taking the exponential,
\begin{align}
    \Omega = e^{-\frac B2}.
\end{align}

In order to verify that this works, let us first write the following commutation relation using the identity $[\Gamma_j,\Gamma_k] = 2\Gamma_j\Gamma_k$ for $k\neq j$,
\begin{align}\begin{aligned}\label{eqn:gamma_b_comm}
    [\Gamma_j, B] &= \frac 12 \omega_{kl} \left [\Gamma_j, \Gamma_k\Gamma_l \right ], \\
    & =   (\omega_{jk}  - \omega_{kj})\Gamma_k. \\
    & = 2\omega_{jk} \Gamma_k.
\end{aligned}
\end{align}
Next, let us consider the following quantity,
\begin{align}
    \chi_j(t) = e^{-t\frac B2} \Gamma_j e^{t\frac B2}.
\end{align}
Using \cref{eqn:gamma_b_comm}, we can show that the following identities hold,
\begin{align}
    \partial_t \chi_j(t) &= \omega_{jk}\chi_k(t),\label{eqn:deriv_chi}\\
     \chi_j(0) &= \Gamma_j.
\end{align}
This allows us to integrate \cref{eqn:deriv_chi}, arriving at the following,
\begin{align}
    \chi_j(t) = \exp(\omega t)_{jk}\Gamma_k.
\end{align}
Finally, setting $t = 1$ and using \cref{eqn:log_R}, we arrive at
\begin{align}
    \chi_j(t) = R_{jk}\Gamma_k.
\end{align}
Finally, inserting the Clifford vector $\slashed{ v}$ into this identity, we find
\begin{align}
    \slashed{ v}' = e^{-\frac 12 B} \slashed { v}e^{ \frac 12 B}.
\end{align}

\subsection{N-dimensional reflections} \label{apx:n_dim_reflections}

Now let us consider representing reflections using the same formalism as above. Again, we have an arbitrary vector $\bm v \in \mathbb R^D$ represented using $\Gamma_j$, as
\begin{align} 
    \slashed { v} & = v_j \Gamma_j.
\end{align}
Let us also consider a unit vector $\hat{\bm n}$, with its corresponding Clifford representation $\hatslashed{ n}$. We may split the vector $\bm v$ into two components, one that is parallel to $\hat{\bm n}$ and one that is perpendicular to $\hat{\bm n}$,
\begin{align}
    \bm v = \bm v_{\parallel} + \bm v_{\perp}, 
\end{align}
where we may write 
\begin{align}
    \bm v_{\parallel} &= (\bm v \cdot \hat {\bm n}) \hat {\bm n}, \\ 
    \bm v_{\perp} & = \bm v - (\bm v \cdot \hat {\bm n}) \hat {\bm n}.
\end{align}
Under the action of the operator $A$, which performs a reflection in the hyperplane perpendicular to $\hat {\bm n}$, we expect that the vector should transform according to
\begin{align} \label{eqn:reflected_v}
    \bm v' = A \bm v = -\bm v_{\parallel} + \bm v_{\perp}.
\end{align}
Now, let us consider 
\begin{align}\begin{aligned}
    \hatslashed{ n}\slashed { v} \hatslashed{ n} &= \hat n_j v_k \hat n_l \Gamma_j\Gamma_k\Gamma_l, \\ 
    & = - v_k \Gamma_k \cdot (\hat n_j\hat n_l \Gamma_j \Gamma_l)   
    +  \hat n_j v_k \hat n_l 2\delta_{jk} \Gamma_l, \\ 
    & = -v_k \Gamma_k + 2 (\bm v \cdot \hat {\bm n})\hat n_l \Gamma_l,
\end{aligned}
\end{align}
where we have used the identity $\hat n_j\hat n_l \Gamma_j \Gamma_l = |\hat {\bm n}|^2 = 1$. Finally, rewriting the above in terms of slashed vectors, we find
\begin{align}\begin{aligned}
    \hatslashed{ n}\slashed { v} \hatslashed{ n} &= - \slashed{ v} + 2(\bm v \cdot \bm n) \hatslashed { n},\\
    & =  \slashed { v}_{\parallel}-\slashed { v}_{\perp}.
\end{aligned}
\end{align}
Finally, comparing with \cref{eqn:reflected_v}, we see that the reflected vector is given by,
\begin{align}
    \hatslashed{ v}' = -\hatslashed{ n}\slashed { v} \hatslashed{ n}.
\end{align}

\section{Almost-commutation and locality}\label{apx:locality}

In this section we study the notion of locality, and the constraints it places on the Hamiltonian $H$ and position operator $X$. Our aim is to show that the condition discussed in \cref{eqn:HX_comm}, where we require that the Hamiltonian \textit{almost commutes} with a position operator $X$, is equivalent to demanding that the Hamiltonian is local. This condition can loosely be interpreted as the absence of long-ranged hoppings. We shall find that bounding the commutator between $H$ and $X$ is equivalent to demanding that the hoppings of $H$ have at least power law scaling with distance.

In this section we shall parametrise $H$ explicitly in real space as
\begin{align}
    H = \sum_{jk} \ketbra {\bm r_j}{\bm r_k} \otimes h_{jk},
\end{align}
where $h_{jk}$ is a hopping element, or on-site term when $j=k$. Here we will treat $h_{jk}$ as a number, however the argument presented is easily extended to matrix-valued $h_{jk}$. The position operator is defined by the relationship $X \ket{\bm r_j} = x_j \ket{\bm r_j}$.


Let us consider the commutator between $X$ and $H$. Since they are both Hermitian operators, their commutator is can be expressed in terms of the Hermitian velocity operator $v$,
\begin{align}
    v = -i[H,X].
\end{align}
Inserting the expression for $H$ gives an explicit expression for the velocity,
\begin{align} \label{eqn:v_explicit}
    v =  \sum_{jk} -i(x_j - x_k)
    \ketbra {\bm r_j}{\bm r_k} \otimes h_{jk}.
\end{align}
The quantity we are interested in is the spectral radius $\norm{v}_2$, defined as the absolute value of the largest eigenvalue of $v$. This satisfies the following expression \cite{Horn1985},
\begin{align}
    \norm{v}_2 = \max_{a, b} |\braket{a|v|b}|,
\end{align}
for arbitrary normalised vectors $a$ and $b$. Thus, inserting position eigenstates into the above expression we find a bound on the individual elements of $v$,
\begin{align}
    |v_{jk}| \leq \norm{v}_2 \, \forall j,k.
\end{align}
Now, inserting \cref{eqn:v_explicit} into the above inequality, we find the following expression
\begin{align}
    |h_{jk}| |x_j - x_k| \leq \norm{[H,X]}_2.
\end{align}
Finally, if we demand that $X$ and $H$ almost commute, placing a bound on the norm of their commutator, 
\begin{align}
    \norm{[H,X]}_2 \leq \mu,
\end{align}
for some chosen value of $\mu$, we find that this is equivalent to imposing the condition on $h_{jk}$,
\begin{align}
    |h_{jk}| \leq \frac \mu {|x_j - x_k|}.
\end{align}
Thus, we see bounding the commutator imposes a power law (or faster) scaling of the couplings with distance.

\end{document}